\documentclass[a4paper,fleqn,usenatbib]{mnras}
\usepackage{mathptmx}
\usepackage[T1]{fontenc}
\usepackage{ae,aecompl}
\usepackage{xcolor}
\usepackage{float}

\usepackage{graphicx}	
\usepackage{amsmath}	
\usepackage{amssymb}	
\usepackage{epstopdf}
\usepackage[normalem]{ulem}
\usepackage{subcaption}
\usepackage{setspace}

\newcommand{\be}{\begin{equation}}
\newcommand{\ee}{\end{equation}}
\newcommand{\ba}{\begin{align}}
\newcommand{\ea}{\end{align}}

\title[Measurement of neutron star mass and radius]{Atmospheric oscillations provide simultaneous measurement of neutron star mass and radius}

\author[D. A. Bollimpalli, M. Wielgus, D. Abarca, W. Klu\'{z}niak]{
	D. A. Bollimpalli,$^{1,2}$\thanks{E-mail: \href{mailto:deepika@camk.edu.pl}{deepika@camk.edu.pl}}
	M. Wielgus,$^{3,1,2}$\thanks{E-mail: \href{mailto:maciek.wielgus@gmail.com}{maciek.wielgus@gmail.com}}
	D. Abarca,$^{1}$\thanks{E-mail: \href{mailto:dabarca@camk.edu.pl}{dabarca@camk.edu.pl}}
	W. Klu\'{z}niak$^{1,2}$\thanks{E-mail: \href{mailto:wlodek@camk.edu.pl}{wlodek@camk.edu.pl}}
	\\
	$^{1}$Nicolaus Copernicus Astronomical Center, ul. Bartycka 18, PL 00-716 Warsaw, Poland\\
	$^{2}$Kavli Institute for Theoretical Physics, University of California, Santa Barbara, CA 93106, USA\\
	$^{3}$Black Hole Initiative, Harvard University, 20 Garden Str, Cambridge, MA 02138, USA\\
}
\date{Accepted XXX. Received YYY; in original form ZZZ}

\pubyear{2018}

\begin{document}
\label{firstpage}
\pagerange{\pageref{firstpage}--\pageref{lastpage}}
\maketitle
\begin{abstract}
Neutron stars with near-Eddington observable luminosities were shown
to harbour levitating atmospheres, suspended above their surfaces. We
report a new method to simultaneously measure the mass and radius of a
neutron star based on oscillations of such atmospheres.  In this paper,
we present an analytic derivation of a family of relativistic,
oscillatory, spherically symmetric eigenmodes of the optically and
geometrically thin levitating atmospheres, including the damping
effects induced by the radiation drag. We discover characteristic
maxima in the frequencies of the damped oscillations and show that from a measurement of the frequency maximum and of the luminosity one can determine the mass and radius of the
neutron star. In addition to the stellar parameters, observation of the variation of the oscillation frequencies with flux would allow us to estimate the stellar luminosity and therefore the distance to the source with an accuracy of a few per cent. We also show that the
ratio of any two undamped eigenfrequencies depends only on the
adiabatic index of the atmosphere, while for the damped eigenfrequencies,
this ratio varies with the luminosity. The damping
coefficient is independent of the mode number of the oscillations. 
Signatures of the dynamics of such atmospheres will be reflected in the
source's X-ray light curves.
 


\end{abstract}





\section{Introduction}
Neutron stars are the most compact non-singular objects observed in the Universe. The theoretical
densities of these objects range from about $10^4~\rm{g}~\rm{cm}^{-3}$ at the surface of the crust \citep{chamel+08} 
to above nuclear density of around $10^{14}~\rm{g}~\rm{cm}^{-3}$ in the core \citep{rhoades+74}. 
The physics governing the equation of state at such large densities becomes uncertain, and a wide
range of equations of state has been proposed, each producing different mass--radius relations,
for neutron stars \citep{arnett+77, Coop1988, cook+94, lattimer+12}.
Accurate measurements of both the mass and radius of neutron stars are necessary for constraining the equation of state, and thus the physics of highly dense material. 
Precise measurements of the mass are possible
for pulsars in binary systems \citep{hulse+75, ozel+16}.
In semidetached low-mass X-ray binaries (LMXBs), stellar masses can only be roughly estimated by the usual methods of determining the orbital parameters \citep{charles+11}.

Measurements of radius are more difficult and they are usually performed for X-ray sources that are not radio pulsars. 
Most methods involve X-ray spectroscopy, for example during thermonuclear bursts, when the redshift  may be inferred by comparing the flux at maximum radius expansion of the burst with that at ``touchdown''
\citep{ Ebi+87, vanParadijs87, damen+90, nattila+16},
or in the quiescent state of LMXBs \citep{BBR+98, marino+18}.
The determined radius usually depends on the mass -- in this sense the spectral method yields simultaneous measurements of both the mass and the radius, although the confidence contours give rather large (banana-shaped) allowed areas in the mass--radius plot \citep[e.g.,][]{bogdanov+16}. To date, sufficiently accurate results have not been obtained, or at least in enough systems, to effectively
measure the equation of state, although constraints have been imposed. 

We have found a new method for an accurate simultaneous measurement of both the mass and the radius based on the expected timing properties of accreting
neutron stars at near-Eddington luminosities, if the distance is known.
Potentially, if a certain type of atmospheric oscillation is detected over a wide range of X-ray fluxes in an X-ray burst source, the distance to the source will also be directly measurable from the timing properties.

Super-Eddington luminous neutron stars are observed in various astrophysical phenomena and it is understood that such high luminosities are powered by the accretion of matter on to the stellar surface or surface nuclear burning. Recent studies have reported highly luminous neutron stars to be such accretors in pulsing ultraluminous X-ray sources NGC 7793 P13,
NGC 5907, NuSTAR J09551+6940.8, NGC 300 ULX1 \citep{bachetti, israel2016, israel2017, carpano+18}. Super-Eddington luminosity has long been known to occur during the outbursts in a few transient X-ray binaries e.g., A0538--66, SMC X-1 and GRO J1744--28 \citep{1982nat, groj, engel}, and during the X-ray bursts. 

Type-I X-ray bursts, are powered by thermonuclear burning of accreted material, during which neutron stars often reach Eddington luminosities \citep{lewin1984, tawara1984, lewin+93}. Spectral-timing analysis of these X-ray bursts shows periodic intensity variations that are termed as burst oscillations \citep{Strohmayer2006}. Such oscillations have been observed to occur during the rise, peak and/or decay phases of the X-ray bursts with typical frequency range within 245--620 Hz \citep{Strohmayer1996,watts2012}. Burst oscillations during the rise phase are usually attributed to the rotational modulation of the  hot spot corresponding to flame propagation on the stellar surface. Observations of burst oscillations from nuclear powered pulsars that emit persistent accretion-powered pulsations have shown that the burst oscillation is correlated with the spin of the neutron star in SAX J108.4--3658 \citep{deepto2003} and 4U 1636--53 \citep{SM2002}. However, the origin of burst oscillations that occur during the decay phase of the X-ray bursts is not yet well understood. Here, we explore another possibility of understanding these frequencies along with a new method to measure the mass and radius of accreting neutron stars in the context of the radial oscillations of ``levitating atmospheres". 

For compact stars with Eddington luminosities, radiation pressure becomes quite significant close to the stellar surface and thus has a strong influence on the dynamics of the ambient material. There is an effect unique to general relativity when examining the forces on the fluid outside a highly luminous star. If we take the fluid to be optically thin, then there exists exactly one radius,~$r_0$, where the gravitational acceleration is balanced by the radiative force of a spherically symmetric star~\citep{Abramowicz1990}. The force balance can be described by an effective potential well centred at $r_0$. Later studies on the motion of test particles in the combined field of gravitation and radiation have shown that the radiation drag has significant effects on the particle dynamics \citep{Bini2009,oh2011,Stahl2012}. A particle orbiting a~highly luminous star has its angular momentum removed by the radiation field due to the Poynting--Robertson effect (or radiation drag effect), causing it to come to rest at this special radius $r_0$. For this reason, this radius is also deemed as the ``Eddington capture sphere", or ECS \citep{Wielgus2012,Stahl2013}. These results generalize to a~shell of optically thin fluid, in which case a~stable levitating atmosphere  suspended around $r_0$ is formed \citep{Wielgus2015}, which is completely supported by the radiation pressure, and not in contact with the star at all. It has been proposed that oscillations of these radiation supported atmospheres could be good candidates to explain the oscillation frequencies observed during Type~I X-ray bursts. \cite{Abarca2016} examined the lowest order, incompressible, radial mode of such atmospheres, and found it to be overdamped by radiation drag. Encouraged by the possibility of higher order modes, \cite{Bollimpalli2017}, performed a~full analysis in a Newtonian framework without radiation drag, finding a~family of eigenmodes with promising eigenfrequencies. For values typical for a~near-Eddington neutron star, the natural frequency is right around 600 Hz. What remains to be done is a~similar analysis in a~general relativistic framework with a proper treatment of radiation drag, in quest for underdamped oscillations. In this paper, we perform such an analysis. 

The paper is outlined as follows. In Section~\ref{Section2}, we give a pedagogical introduction to the radius of ``force balance'' and to the fundamental frequencies. We shall then discuss in Section~\ref{Section3}, the necessary equilibrium conditions for levitating atmospheres and derive a family of relativistic eigenmodes and eigenfrequencies of the radial oscillations. Damping of oscillations owing to radiation drag effect is investigated and an analytical expression for the damping coefficient is also found. In Section~\ref{Section4}, we demonstrate how a new method to measure the mass and radius of neutron stars emerges from a unique maximum frequency treated as a function of stellar luminosity, mass and radius, as well as the oscillation mode.
In Section~\ref{Section5}, we show how observing changes in the frequency over a wide range of X-ray fluxes would allow a determination of the distance to the source as well.  We conclude the paper by a discussion (Section~\ref{SectDisc}) and a summary of the results (Section~\ref{SectConc}). 

\section{Fundamentals of levitating atmosphere oscillations}
\label{Section2}
Radial oscillations of levitating atmospheres have been previously investigated for special cases \citep{Abarca2016, Bollimpalli2017}, and the current paper aims to present a general study of these oscillations, including radiation drag in the relativistic regime. Before turning to rigorous calculations (Section~\ref{Section3} and following), we present a heuristic derivation of the fundamental mode frequency of the levitating atmosphere.

The usual Eddington luminosity derivation assumes that both the radiative flux and the force of gravity are inversely proportional to the square of the radius, $r$, of a spherical surface concentric with the star. In general relativity this is no longer the case -- because it is redshifted, the radiative flux decays with distance more rapidly than the gravitational acceleration -- and (in the spherically symmetric case) there will be only one particular radius, $r_0$, at which the radiative force on the electron of a stationary hydrogen atom can balance the gravitational acceleration of the proton. At lower radii, radiation force dominates over gravity and so (in Newtonian language) the net force is directed away from the star, towards the ECS at $r_0$. At larger radii, beyond $r_0$, gravitational attraction dominates the radiation force and so the net force is  directed towards the star, i.e., again towards the ECS. A stable, optically thin fluid shell may therefore be positioned at $r\approx r_0$. In hydrostatic equilibrium the net force (``effective gravity'') is balanced by the fluid pressure gradient -- thus at $r_0$, where the effective gravity vanishes, the pressure is at its maximum. 

If the radial extent of such a fluid shell is much smaller than $r_0$ (``thin atmosphere limit''), its fundamental mode of radial oscillations corresponds to uniform radial displacement by a distance $\delta s(r,t) = \delta s(t)=(1+z)\delta r(t)$ -- the redshift factor ($1+z$) accounting for the difference between proper and coordinate distance -- with the shell suffering a restoring force per unit mass of  $-\omega^2 \delta s(t)$, resulting in harmonic radial oscillations at the (locally observed) frequency
$\omega =\sqrt{({\rm d}\, f_g/{\rm d}\, r)( \delta r/\delta s)}$, with $f_g(r)$ being the value of the net force per unit mass at $r$, and the derivative evaluated at $r_0$. 
\footnote{The higher radial modes are more difficult to derive, as pressure variations play a role in determining the oscillatory motion. For example, the next higher mode is a breathing mode in which the oscillatory motion can be described as expansion away and contraction towards the ECS, with the (only) node at the ECS \citep{Bollimpalli2017}.}

To derive $r_0$ and $\omega$, we note that for an observer measuring the stellar luminosity, both the frequency of the photons and the rate of the arrival of photons are gravitationally redshifted, introducing two redshift factors, i.e., $(1+z)^{-2}$. Therefore, for a static, spherically symmetric field around a star with mass $M$ and radius $r_*$, a local observer at radius $r$ measures the stellar luminosity $L(r)$ to be decreasing with the radial coordinate as
\begin{equation}
L(r) = L_{\infty}\left(1-\dfrac{2GM}{rc^2}\right)^{-1} = L_* \left(1-\dfrac{2GM}{r_*c^2}\right)\left(1-\dfrac{2GM}{rc^2}\right)^{-1},
\label{lstar}
\end{equation}
where $L_{\infty}$ is the luminosity measured by a distant observer $(r~\rightarrow~\infty)$,  $L_*$ would be measured at the stellar surface,
and   $1+z = \left[1-{2GM}/\left({rc^2}\right)\right]^{-1/2}$ is the redshift factor.

Now, the gravitational acceleration of a static observer at radius $r$ is given by $GMr^{-2}(1+z)$. Therefore, at the radius $r_0$, where the gravitational acceleration and the radiative force of the star balance each other at local luminosity $L(r_0)$, one has
\begin{equation}
  L_{\rm Edd}\left(1-\dfrac{2GM}{r_0c^2}\right)^{-1/2} = L(r_0) =
  L_{\infty}\left(1-\dfrac{2GM}{r_0c^2}\right)^{-1},
\label{ledd}
\end{equation}
where $L_{\rm Edd}=4 {\mathrm \pi} c G M/\kappa$ is the usual expression for the Eddington luminosity, with the opacity given by $\kappa$. Defining
\be
\lambda = L_\infty/L_\mathrm{\rm Edd}
\label{lambda}
\ee
for our convenience,
we immediately obtain
\be
\lambda =\left(1-\frac{2GM}{r_0 c^2}\right)^{1/2}
\label{lambda2}
\ee
from eq.~(\ref{ledd}), and the radius of equilibrium follows
as \citep{Abramowicz1990,Stahl2013}
\begin{equation}
r_0 = \frac{2GM}{c^2(1 - \lambda^2)}.
\label{r_ecs}
\end{equation}
For a fixed value of $\lambda$, the radius scales directly with mass, as is usual in general relativity.

The radiative force per unit mass is
  $\kappa L(r)/(4{\mathrm \pi} c r^2)=\lambda(GM/r^2)[1- 2GM/(rc^2)]^{-1}$,
  where we used the first equality in eq.~(\ref{lstar}) and the definition of
  $\lambda$ and $L_\mathrm{\rm Edd}$.
  The locally observed eigenfrequency of the fundamental mode, $\omega$, and its counterpart observed at infinity,  $\omega_r=\omega/(1+z)$, can now easily be obtained by expanding to first order in $r-r_0$ the expression for the effective gravity (difference between the radiative force per unit mass and the acceleration of gravity)
\begin{align*}
& -\frac{GM}{r^2}\left[ \left(1-\frac{2GM}{rc^2} \right)^{-1/2} -\left(1-\frac{2GM}{rc^2} \right)^{-1} \left(1-\frac{2GM}{r_0 c^2}\right)^{1/2} \right]\\
 &\approx
-\left(\frac{GM}{r_0^2c}\right)^2(1+z)^2 \delta s\equiv- \omega^2\delta s.
\end{align*}
Finally \citep{Abarca2016}, 
\be
\omega_r=\frac{GM}{r_0^2c}=\frac{c^3(1-\lambda^2)^2}{4GM}.
\label{eq:FreqDavid}
\ee

Note the $1/M$ general relativistic (GR) scaling of the frequency:
\be
\omega_r=\frac{c^3}{GM}\left(\frac{GM}{r_0 c^2}\right)^2,
\label{eq:GR}
\ee
the expressions in parentheses being constant for fixed $\lambda$ in eqs.~(\ref{r_ecs})~--~(\ref{eq:GR}).
As the observable frequency depends on the observed flux and the neutron star mass, it is then possible to determine the stellar mass from the two observables,  $\omega_r$ and $L_{\infty}$. In the following sections, we will see that when radiation drag is included in the calculation, the frequency has a maximum corresponding to a location of the levitating atmosphere that is quite close to the stellar surface, $r_0\approx r_*$. As $r_0$ is also determined from the two observables, this allows a determination of the stellar radius as well. More precise statements can be found in the remainder of the paper.

\section{Levitating atmospheres: Hydrostatic Equilibrium and Radial Oscillations}
\label{Section3}
In this section, we present the full general relativistic calculations for radial perturbations of an optically thin levitating atmosphere. We adopt the natural system of units in which $G=1=c$. We assume a static spherically symmetric space--time for the background solution, thereby the line element for a Schwarzschild metric with signature ($-$,+,+,+) is given by
\be
ds^2 = g_{ij}dx^i dx^j =  -B dt^2 + B^{-1} dr^2 + r^2(d \theta^2 +\sin^2 \theta d\phi^2),
\label{mertic}
\ee
where we have introduced
\be 
B \equiv 1 - \frac{2M}{r} = -g_{tt} = g^{rr} .
\label{B}
\ee
We assume a perfect fluid with stress energy tensor $T^{\mu \nu} = (\epsilon+p+\rho) u^{\mu}u^{\nu} +pg^{\mu \nu}$, where $u^{\mu}$ is the four-velocity, $\epsilon$ is the internal energy, and $\rho$, $p$ are rest mass density and pressure of the flow, respectively.  For the relevant gas temperatures, $T<10^{10}\,K$, it is safe to assume that $\rho \gg p+\epsilon$. We keep this simplifying assumption throughout the work. 

\begin{figure}
\includegraphics[scale=0.41]{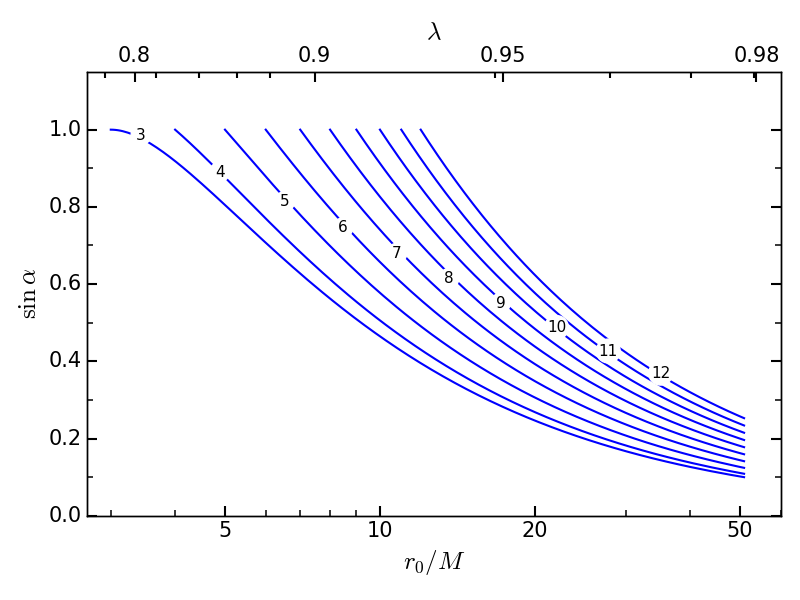}
\caption{The sine of the viewing angle as a function of the ECS location, plotted for a range of stellar radii $r_*=$ $3$ to $12M$ (from left to right).}
\label{fig:sina}
\end{figure}

We start with the governing equations of fluid dynamics; the continuity equation reads
\be 
\nabla_\mu \left( \rho u^\mu \right) = 0,
\ee
where $\nabla_\mu$ is the covariant derivative. The conservation of energy--momentum is given by 
\be
\nabla_{\mu}T^{\mu \nu} = G^{\nu},
\ee
where $G^{\nu}$ is the radiation four-force density \citep{Mihalas84}. For electron-scattering-dominated regions, $G^{\nu} = \kappa \rho F^{\nu}$, where $\kappa$ is the scattering opacity and $F^{\nu}$ is the radiation flux. Using the projection tensor $h^{\nu}_{\ \mu}=\delta^{\nu}_{\ \mu}+u^\nu u_\mu$, this quantity can be determined through the radiation stress--energy tensor, $R^{\mu \nu}$, as $F^{\nu} = h^{\nu}_{\ \lambda}R^{\mu \lambda}u_{\mu} $. Under the optically thin limit, all components of $R^{\mu \nu}$ for an isotropically radiating star were first presented in \citet{Abramowicz1990} and we adopt the same here. The relevant components of the radiation stress--energy tensor are
\begin{eqnarray}
\label{rad_tensor_comps1}
R^{tt} &=& \dfrac{L(r)}{2 \pi r_*^2}\left(1-\dfrac{2M}{r_*}\right)\left(1-\dfrac{2M}{r}\right)^{-2}(1-\cos \alpha) , \\
R^{tr} &=& \dfrac{L(r)}{4 \pi r_*^2}\left(1-\dfrac{2M}{r_*}\right)\left(1-\dfrac{2M}{r}\right)^{-1} \sin^2 \alpha ,  \\
R^{rr} &=&  \dfrac{L(r)}{6 \pi r_*^2}\left(1-\dfrac{2M}{r_*}\right)(1-\cos^3 \alpha) .
\label{rad_tensor_comps3}
\end{eqnarray}
where $r_*$ is the radius of the star, and $\alpha$ is the viewing angle, which is the maximum polar angle up to which an observer at radius $r$ can see the photons from the stellar surface. For simplicity, we assumed that the stellar radius is beyond the photon orbit, $r_*>3M$, which gives\footnote{Refer to \citet{Abramowicz1990} for details.}
\be 
\sin \alpha = \dfrac{r_*}{r}\left(1-\frac{2M}{r}\right)^{1/2}\left(1-\frac{2M}{r_*}\right)^{-1/2}.
\label{viewang}
\ee
The dependence of $\sin\alpha$ on $r_0$ is shown in Figure~\ref{fig:sina} for several stellar radii.

The radial component of the relativistic Euler equation for an optically thin gas subject to the radiation field becomes
\be 
u^\mu \nabla_\mu u^r + \left( g^{\mu r} + u^\mu u^r \right) \frac{\nabla_\mu p}{\rho} + \kappa u_{\mu}\left(R^{r\mu} +u^ru_{\nu}R^{\mu \nu}\right) = 0 \ .
\label{eq:Euler}
\ee

We have assumed a uniformly radiating star, which may be a good description of a neutron star undergoing an X-ray burst.
For an accreting and rotating neutron star, a rotating, radiating and levitating belt in the boundary layer has been shown to form due to the balance of gravitation, centrifugal and radiation forces; as the accretion rate approaches the near-Eddington luminosities, this belt extends from the equatorial plane up to the pole, allowing the entire surface of the neutron star to radiate \citep{Inogamov1999}. Thus, the spherically symmetric envelope may be a useful approximation even for a rotating neutron star accreting through a disc.

\subsection{Geometrically thin background solution}
Initially, the atmosphere is in hydrostatic equilibrium with a normalized four-velocity, $u^t$, given by $(B^{-1/2},0,0,0)$. Following equation~(\ref{eq:Euler}), the hydrostatic equilibrium condition for a spherically symmetric and optically thin levitating atmosphere around a uniformly radiating star is 
\citep{Wielgus2015} 
\be
\dfrac{1}{\rho} \dfrac{d p}{d r} = -\dfrac{M}{r^2 B}\left(1-\dfrac{\lambda}{B^{1/2}} \right).
\label{he}
\ee
Note that the location of the pressure maximum from the above equation, at $\lambda=B_0^{1/2}$, corresponds to the location of the ECS, $r_0$, see also equation~(\ref{r_ecs}),
\begin{equation}
\label{ECS}
r_0 = \dfrac{2M}{1-\lambda^2}.
\end{equation}
Quantities with subscript $0$ are evaluated at the ECS, e.g., $B_0\equiv B(r_0)=1-2M/r_0$.
 Analytical solutions of an optically and geometrically thin polytropic atmosphere show that the density and pressure have Gaussian-like profiles \citep{Wielgus2015} with the maximum located at $r_0$. Here, we present such solutions in a form usable for the calculation of normal modes. The procedure is quite similar to the one employed in the task of finding normal modes of oscillating tori (see, e.g. \citet{Blaes2006}. The main difference is that in the currently considered case the radiation force, rather than the centrifugal force, balances gravity in the background solution. The simplification is that the current problem has a~spherical symmetry; hence, we obtain ordinary (and not partial) differential equations. Denoting $\mathcal{E} = -u_t=B^{1/2}$, we define an effective potential $\mathcal{U}_{\rm eff}$ such that
\be 
\frac{1}{\rho}\frac{d p}{d r} = -\frac{\mathcal{E}^2}{2} \frac{d \mathcal{U}_{\rm eff}}{d r} \ .
\ee
Using equation~(\ref{he}) and integrating, we find
\be 
\mathcal{U}_{\rm eff} = -B^{-1} + \frac{2 \lambda}{3}B^{-3/2} \ .
\ee
Casting the equation of hydrostatic equilibrium as
\be 
\frac{1}{\rho}\frac{d p}{d r} + \frac{\mathcal{E}_0^2}{2}  \frac{d \mathcal{U}_{\rm eff} }{dr} = - \frac{1}{2} \left(\mathcal{E}^2 - \mathcal{E}_0^2 \right) \frac{d \mathcal{U}_{\rm eff}}{d r}\ , 
\label{hydro_eq}
\ee
allows us to express the right-hand side of the equation as a gradient of a scalar function, $ \psi$.  For a polytropic fluid with an adiabatic index $n$, $p \propto \rho^{(1+1/n)}$, we may integrate equation~(\ref{hydro_eq}) to find the Bernoulli equation in the following form:
\be 
(1+n) \frac{p}{\rho} +  \frac{1}{2} \mathcal{E}_0^2 \mathcal{U}_{\rm eff} + \psi  = \text{const} .
\ee
After evaluating the constant at the ECS and some reordering, this integral condition can be cast as
\be
 \frac{p}{\rho} = \frac{p_0}{\rho_0} \left \{ 1- \frac{1}{ n c_{{\rm s},0}^2}\left[ \frac{\mathcal{E}_0^2}{2} \left( \mathcal{U}_{\rm eff} -  \mathcal{U}_{\rm eff, 0} \right)+ \psi - \psi_0 \right] \right \} \ ,
\label{eq:BackNoApprox}
\ee
with the speed of sound (squared) $c_s^2=dp/d\rho$.

So far we did not explicitly assume that the background solution should be geometrically thin. We make this simplification now, by approximating the right-hand side of equation~(\ref{eq:BackNoApprox}) with the second-order Taylor expansion around $r_0$, noticing that
\be 
\left. \frac{d \mathcal{U}_{\rm eff}}{dr} \right|_{r_0} = \left. \frac{d \psi}{dr} \right|_{r_0} = \left. \frac{d^2 \psi}{dr^2} \right|_{r_0} = 0 \ .
\ee
The approximation holds for atmospheres that are sufficiently geometrically thin to neglect the gradient of the potential, i.e., the radial variation of space--time curvature and the radiation field, in comparison with the gradient of pressure. Hence, it is acceptable for atmospheres with thickness much less than $r_0$. The following approximate formula is obtained:
\be
 \frac{p}{\rho} \approx \frac{p_0}{\rho_0} \left [ 1- \frac{1}{ n c_{{\rm s},0}^2}\frac{\mathcal{E}_0^2}{2} \frac{d^2 \mathcal{U}_{\rm eff}}{dr^2} \frac{(r-r_0)^2}{2}
  \right] \equiv \frac{p_0}{\rho_0} f .
  \label{eq:BackApprox0}
  \ee
Seeing how
\be
       \frac{-g_{tt}(r_0) \mathcal{E}_0^2}{2 g_{rr}(r_0)} \left. \frac{d^2 \mathcal{U}_{\rm eff}}{dr^2} \right|_{r_0} = \frac{\lambda^6}{2} \left. \frac{d^2 \mathcal{U}_{\rm eff}}{dr^2} \right|_{r_0} = \left( \frac{M}{r_0^2} \right)^2 \ , 
\label{omegar_20}
  \ee
this result can be cast in a more compact form
using the following substitutions:
\begin{align} 
   \omega_r^2 &=
  \left( \frac{M}{r_0^2} \right)^2 \ , & 
\label{omegar_20}\\
 \beta &= \frac{\sqrt{2 n} c_{{\rm s},0} \lambda}{r_0} \ , & \\
 x &= \frac{r - r_0}{r_0} \sqrt{g_{rr}}  \ , & \\
 \eta &= \frac{\omega_r}{\beta}x \ , & \label{eta_def}
\end{align}
with a simple interpretation: $\omega_r$ corresponds to the fundamental oscillation angular frequency, already familiar from equation~(\ref{eq:FreqDavid}), parameter $\beta$ controls the geometric thickness of the atmosphere, $x$ and $\eta$ correspond to a~convenient scaling of the radial coordinate.
Thus, equation~(\ref{eq:BackApprox0}) for the radial profile of pressure may be rewritten simply as
\be
\frac{p}{\rho} \approx \frac{p_0}{\rho_0} \left(1 - \frac{\omega_r^2}{\beta^2} x^2 \right) = \frac{p_0}{\rho_0} (1 - \eta^2) = \frac{p_0}{\rho_0} f .
\label{eq:ThinAtmosphApprox}
\ee

\subsection{Perturbations of the geometrically thin solution}

We assume coherent, spherically symmetric perturbations of physical variable $X$ in the form of
\be 
\widehat{\delta X}(r,t) = \delta X (r) \exp(-i \omega t) \ 
\ee
and linearly perturb the relevant hydrodynamic equations. The first thing to notice is that $\delta u^t = 0$ is necessary to fulfil the perturbed four-velocity norm, as $u_r = 0$ in
\be 
u_r \delta u^r + u_t \delta u^t = 0 .
\ee
The linearly perturbed version of the continuity equation takes the following form
\be 
-i \omega B^{-1/2}\frac{\delta \rho}{\rho} + \frac{d}{dr} \delta u^r + \frac{1}{\sqrt{-g} \rho} \frac{d  (\sqrt{-g} \rho)}{d r} \delta u^r = 0 \ .
\label{eq:ContinPert}
\ee
The perturbed radial Euler equation yields
\be
-i \omega u^t \delta u^r + g^{rr} \frac{d}{d r} \left(\dfrac{\delta p}{\rho}\right)  + \delta \mathcal{D} = 0 \ ,
\label{eq:EulerPert}
\ee
where $\delta \mathcal{D} = \kappa \left( u_{t}^2 R^{tt}+g_{rr} R^{rr} \right) \delta u^r$ relates to the radiation drag acting against the radial motion of the particle and can be rewritten as
\be
\delta \mathcal{D} = \frac{2}{3 }\dfrac{M\left(1-2M/r_*\right)}{\lambda^3 r_*^2 }(1 - \cos \alpha) (\cos^2 \alpha + \cos \alpha + 4)\delta u^r \equiv \chi \delta u^r,
\ee
cf. equations~(\ref{rad_tensor_comps1})--(\ref{eq:Euler}).
Although $\chi$ has a radial dependence through the cosine and sine functions of the viewing angle, in the geometrically thin limit $r \sim r_0$, hence $\chi$ will be taken to be constant for given stellar parameters and stellar luminosity.

Following \citet{Ipser1992} and \citet{Abramowicz2006}, we define a new variable,
\be 
W = -\frac{\delta p}{u^t \rho} \ .
\ee
From the above definition, and equation~(\ref{eq:EulerPert}), it follows that in the geometrically thin limit
\be 
\delta u^r =  i\dfrac{B}{\omega+i \chi \lambda}\dfrac{{\rm d} W}{{\rm d} r}.
\label{del_ur}
\ee
Substituting $W$ in equation~(\ref{eq:ContinPert}) and combining with equation~(\ref{del_ur}) gives a~single second-order ordinary differential equation. In the geometrically thin atmosphere limit ($\beta \rightarrow 0$), this equation simplifies to 
\be 
f\frac{d^2W}{d \eta^2} -2n\eta \frac{d W}{d \eta}+\dfrac{2n\omega^2}{\omega_r^2}\left(1+i\dfrac{\chi \lambda}{\omega}\right)W = 0
\label{Diff_eq_final}
\ee
with $\omega_r^2$ given by equation~(\ref{omegar_20}).
Note that equation~(\ref{Diff_eq_final})
  can be cast into a convenient dimensionless form
 \be 
 (1-\eta^2) \frac{d^2W}{d \eta^2} -2n\eta \frac{d W}{d \eta}+
 2n\left(\sigma^2 +2i\gamma\sigma\right)W = 0,
\label{dimless}
\ee
with
  \begin{align}
\sigma &\equiv \omega/\omega_r \ ,\\
  \gamma&\equiv\chi \lambda/(2\omega_r)
  =(4-\cos^3 \alpha -3 \cos \alpha)/(3\sin^2\alpha),
\label{chi}
  \end{align}
  and with the specific viewing angle given by
  \be 
   \sin\alpha = \frac{r_*}{r_0}\dfrac{\left(1-2M/r_0\right)^{1/2}}{\left(1-2M/r_*\right)^{1/2}}.
\ee
As we will see, $\tau=1/(\omega_r\gamma)$ is the decay constant (damping time scale) of the oscillatory solutions, so it is important to compare the value of the damping coefficient $\gamma$ to the values $\sigma$ of the (dimensionless) oscillation frequencies for the  undamped solutions.

It is easy to show that outside the photon orbit, i.e., for  $r_0\ge 3M$, the viewing angle decreases monotonically with $\lambda$ (or with the ECS radius $r_0$), and so does the damping coefficient $\gamma$. The maximum value of $\gamma=4/3$ is attained at the stellar radius  $r_0 = r_*$, and $\gamma\rightarrow 1$ as $r_0\rightarrow \infty$, $\alpha\rightarrow 0$, $\lambda\rightarrow 1$. Thus, the radiative drag is always present in the problem at hand, with $4/3\ge \gamma>1$. None the less, it is instructive to consider a simplified problem in which radiative drag is artificially neglected.

\subsection{Undamped oscillations}

Let us now consider the artificial problem of undamped oscillations. This is obtained by blithely neglecting the damping term in equation~(\ref{dimless}), i.e., putting $\gamma=0$, which leads to the Gegenbauer equation,
\be 
(1-\eta^2) \frac{d^2W}{d \eta^2} -2n\eta \frac{d W}{d \eta}+2n\sigma^2 W = 0,
\label{real1}
\ee
The corresponding solutions represent the undamped oscillations with the relativistic eigenfrequencies observed at infinity given by\footnote{Which can also be identified directly from equation~(\ref{freq_damp}) with the damping coefficient set to zero.}
\be
\omega_k^2 = \omega_{r}^2 \sigma_k^2,
\label{undamp}
\ee
where
\be
\sigma_k^2 = \dfrac{k(k+2n-1)}{2n}.
\ee

The relativistic frequencies differ from the Newtonian ones \citep{Bollimpalli2017} by a redshift related factor of $\lambda^2$, while the relativistic eigenmodes are similar to the Newtonian eigenmodes. Nevertheless $\sigma_k$, the ratio of normal mode frequencies to the fundamental mode frequency, remains the same for both the relativistic and non-relativistic cases. The ratio of frequencies of the second (``breathing") mode and the fundamental mode is $\sigma_2 = \sqrt{2+1/n}$. 

\begin{figure}
	\includegraphics[scale=0.43]{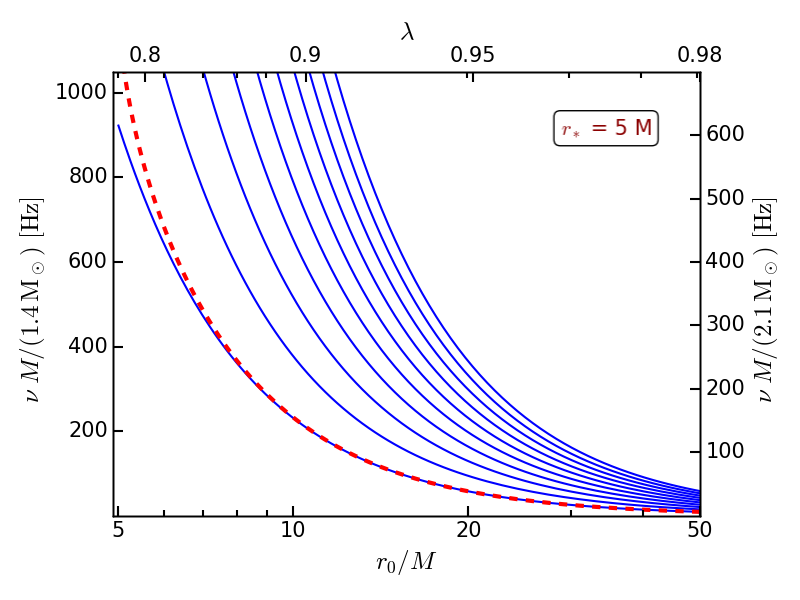}
        \caption{Frequencies of the ten first normal modes of undamped oscillations of the thin atmospheres as function of the atmosphere location. Frequencies scaled to fiducial masses $1.4 \,{\rm M}_{\odot}$ and $2.1\,{\rm M}_{\odot}$ are shown on the left- and right-hand side vertical axes respectively. Here, we assumed the stellar radius $r_*=\,5M$. The dashed (red) line represents the damping rate, scaled in the same way, $\gamma\omega_r\cdot M/(2\pi M_{\mathrm{fiducial}})$. }
    \label{fig:undamp}
\end{figure}

The eigenfrequencies grow with the mode number; the first two are $\sigma_1=1<\gamma$ and $\sigma_2^2={2+1/n}>2>\gamma^2$. Thus, in the full problem, which includes radiative damping, the radial mode ($k=1$) is overdamped \citep{Abarca2016}, but all the higher modes ($k\ge 2$) have an oscillatory character (are underdamped). This is also clear when we plot the frequencies, $\nu = \omega_k(r_0)/(2\pi)$, of the first ten modes of undamped oscillations (left to right), as a function of Eddington parameter or atmosphere location (Figure~\ref{fig:undamp}).

The frequencies are plotted in Hz to facilitate comparison with observations. The values on the left vertical axis correspond to $M = 1.4\, {\rm M}_{\odot}$. For a different mass, $M$, the frequency at a given $r_0$ and a given mode can be obtained with the scaling factor $M/(1.4 {\rm M}_{\odot})$ shown on the left vertical axis. Frequencies computed for $2.1 {\rm M}_{\odot}$ are shown with similar scaling on the right vertical axis. The radius in units of the stellar mass has to be assumed a priori to compute these frequencies, we took $r_* = 5\,M$. For a given $\lambda$ or $r_0$, higher number modes have larger frequencies. The further the atmosphere is located from the stellar radius, the lower the frequencies get as the magnitude of the restoring forces decreases with the radius. This radial dependence is fully accounted for by the functional form of $\omega_r$, given in  equation~(\ref{omegar_20}). We note that there is a range of luminosities for which the frequencies of these oscillations fall in the range of 300--600 Hz, typical for observed frequencies of X-ray burst oscillations. 


\subsection{Damped oscillations}
\label{damped_oscill_sec}
We now turn to the full problem including the damping effects of the ever-present radiation drag in radiation-supported atmospheres.

It is easiest to solve equation~(\ref{dimless}) by considering the real ($\sigma_{\rm R}$) and imaginary ($\sigma_{\rm I}$) parts of $\sigma$ separately,
\be 
\sigma = \sigma_{\rm R} + i \sigma_{\rm I},
\label{omega_ri}
\ee
and the corresponding relation $\omega=\omega_{\rm R}+i\omega_{\rm I}$, with $\omega_{\rm R}=\omega_r\sigma_{\rm R}$ and $\omega_{\rm I}=\omega_r\sigma_{\rm I}$. Substituting complex $\sigma$ into equation~(\ref{dimless}) results in a differential equation with real and imaginary parts that are separable and can be solved simultaneously. The real part is an eigenvalue problem in the form of a Gegenbauer differential equation,
\be 
(1-\eta^2) \frac{d^2W}{d \eta^2} -2n\eta \frac{d W}{d \eta}+2n\left({\sigma_{\rm R}^2-\sigma_{\rm I}^2-2\gamma\sigma_{\rm I}}\right)W = 0,
\label{real}
\ee
while the imaginary part yields
\be 
i 4n\sigma_{\rm R}\left(\sigma_{\rm I}+\gamma\right)W = 0.
\label{imaginary}
\ee

A particularly simple solution corresponds to the overdamped case of purely imaginary $\sigma$. The eigenvalues are given by,
$-\sigma_{\rm I}^2-2\gamma\sigma_{\rm I}=\sigma_k^2$, giving,
  \be
  \sigma_{\rm I}=-\gamma \mp \sqrt{\gamma^2-\sigma_k^2}.
  \ee
In fact, as follows from the discussion above, this solution only occurs for $k=1$, for which $\sigma_k=1$, corresponding to the imaginary frequency of the overdamped solution of
  \be
  \omega=i\omega_{\rm\, OD},
  \ee
  with
  \be
  \omega_{\rm\, OD}=-\omega_r\left[\gamma \pm \sqrt{\gamma^2-1} \right].
\ee
The time dependence of this decaying solution is $\exp(-i\omega\, t)=\exp(\omega_{\rm\, OD}\, t)$. 
\begin{figure}
	\includegraphics[scale=0.43]{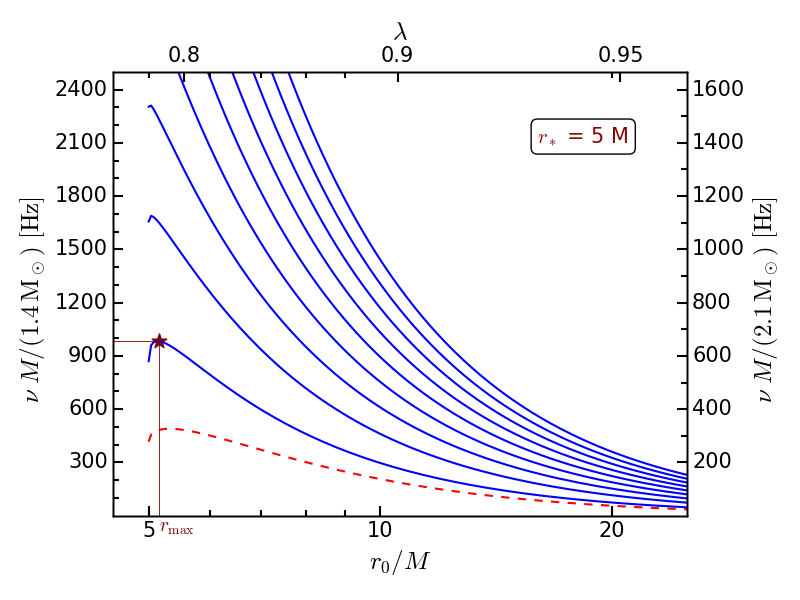}
    \caption{Frequencies of the ten first normal modes of damped oscillations of the thin atmospheres as function of the atmosphere location. For the overdamped fundamental mode, $|\omega_{\rm\, OD}|/(2 \pi)$ is shown by dashed (red) line. Assumed stellar radius is  $r_*=5M$. }
\label{fig:eigefr}
\end{figure}

In the general case, equation~(\ref{real}) can be solved with a proper boundary condition of pressure vanishing on the atmosphere boundary. The corresponding eigenmodes are represented by the Gegenbauer polynomials $C^{\alpha}_k(\eta)$ with $\alpha = n - \frac{1}{2}$ and $k=0,1,2,3,...$ which can be calculated with a recursive formula \citep{Koornwider2010} 
\begin{align}
C^{\alpha}_0(\eta) &= 1 \nonumber \ , \\
C^{\alpha}_1(\eta) &= 2 \alpha \eta \ , \nonumber \\
k C^{\alpha}_{k}(\eta) &= 2 \eta \left( k - 1 + \alpha \right) C^{\alpha}_{k-1}(\eta) - \left( k - 2 + 2 \alpha \right)C^{\alpha}_{k-2}(\eta) \ .
\label{eq:Gegenbauer}
\end{align}
They form a complete set of modes, that is, any spherically symmetric oscillation of the levitating atmospheres is necessarily a~combination of the eigenmodes described by the equation~(\ref{eq:Gegenbauer}). For example, the eigenfunctions of the second mode ($k = 2$) and the third mode ($k=3$), take the form of $\left(n-\dfrac{1}{2}\right)[\eta^2(2n+1)-1]$ and $\dfrac{\eta}{6}(4n^2-1)[(2n+3)\eta^2-3]$, respectively. \footnote{As a reminder, $n$ is the adiabatic index and $\eta$ is a scaled radial coordinate, (see eq.~\ref{eta_def}).} The above imply a non-vanishing eigenfunction W, and for $k>1$ also $\sigma_{\rm R} \neq 0$, so equation~(\ref{imaginary}) simply gives the damping rate
\be
\omega_{\rm I} = -\omega_{r} \gamma.
\label{damp_coeff}
\ee
Using this, the corresponding eigenfrequencies of the oscillations are computed from the Gegenbauer relation,
\be 
\omega_{\rm R}^2 = \omega_{r}^2 \left(\sigma_k^2 -\gamma^2 \right)
 =  \omega_{r}^2 \left[\frac{k(k+2n -1)}{2n} -\gamma^2 \right],
\label{freq_damp}
\ee
where $k=2,3,4...$.
Note that the time dependence of the solution is $\exp(-i\omega_{\rm R}\, t-\omega_{r} \gamma\, t)$, so $1/(\omega_{r} \gamma)$ is the damping time. The damping coefficient $\gamma$ has been defined in equation~(\ref{chi}) and the fundamental frequency $\omega_{r}$ in equation~(\ref{omegar_20}).

The frequency $\omega_{\rm R}^2$ is a product of the fundamental frequency $\omega_r$ squared, which is a rapidly and monotonically decreasing function of the ECS radius $r_0$, and of a monotonically increasing function of the same variable, $\sigma_k^2 -\gamma^2(r_0)$. The product has a maximum, close to the stellar surface, even though the damping coefficient $\gamma$  of equation~(\ref{chi}) is a slowly varying function of the viewing angle, and $\sigma_k$ is a constant. This is because close to the stellar surface, the angle $\alpha$ itself varies rapidly with $r_0$. Indeed,
\be
 \frac{d\gamma}{d r_0}
 =\frac{d\gamma}{d\alpha}\times\frac{1}{\cos\alpha}\times\frac{d\sin\alpha}{d r_0},
 \label{blowsup}
 \ee
 and the second factor dominates the radial derivative of  $\omega_{\rm R}^2$, as  $\cos\alpha\rightarrow 0$  at $r_0\rightarrow r_*$, where $\alpha\rightarrow{\mathrm \pi}/2$, while the other terms are regular everywhere:
 \be
  \frac{d\gamma}{d\alpha}
  = \frac{(1-\cos\alpha)^3(3+\cos\alpha)}{3\sin^3\alpha},
  \ee
  \be
  \frac{d\sin\alpha}{d r_0}
  = - \frac{\sin\alpha}{r_0}\frac{(1-3M/r_0)}{(1-2M/r_0)}.
  \ee
  Thus, close to the stellar surface $\omega_{\rm R}$ is a growing function of $r_0$, while at larger radii, over most of its range, $\omega_{\rm R}$ decreases with distance to the star. Therefore, for any mode higher than the fundamental ($k>1$), a maximum of  the atmospheric oscillation frequency occurs close to the stellar surface. The fundamental mode ($k=1$) is overdamped, and hence the maximum occurs in the inverse decay constant  $|\omega_{\rm\, OD}|$, as can be seen in Figure~\ref{fig:eigefr}.

Figure~\ref{fig:eigefr} presents the frequencies $\omega_{\rm R}(r_0)/(2\pi$) of the initial few lowest modes of the underdamped oscillations in an increasing order of mode number from left to right. One may immediately notice from the figure that the frequencies of the modes have now decreased slightly due to damping. For the fundamental mode, we plot the magnitude, $|\omega_{\rm\, OD}|/(2 \pi)$, shown by the thin dashed line in the plot. The overdamped solutions are not of much astrophysical interest, so in further discussions we shall only consider the second and higher modes, as they allow for oscillations. 

Observed oscillations in the X-ray bursts occur mainly during the rise and decay phase of the outburst \citep{Muno2001}. Interestingly, even with the damping, the observed range of 300--600~Hz still falls within the frequency range of these mode oscillations. Oscillations during the outburst decay phase are found to have increasing frequency with time \citep{Strohmayer2006}. Eq.~(\ref{freq_damp}) and Figure~\ref{fig:eigefr} show that the frequencies increase with decreasing luminosity (as long as the frequency maximum is not reached), which is in accord with the observed oscillations, since luminosity decreases with time during the decay phase of the outbursts, as the name suggests. However, the magnitude of the frequency increase in the optically thin model exceeds the observed changes.


In the following section, we present a method of simultaneous mass and radius measurements from the frequency maximum of the damped oscillations.

\begin{figure}
\includegraphics[scale=0.43]{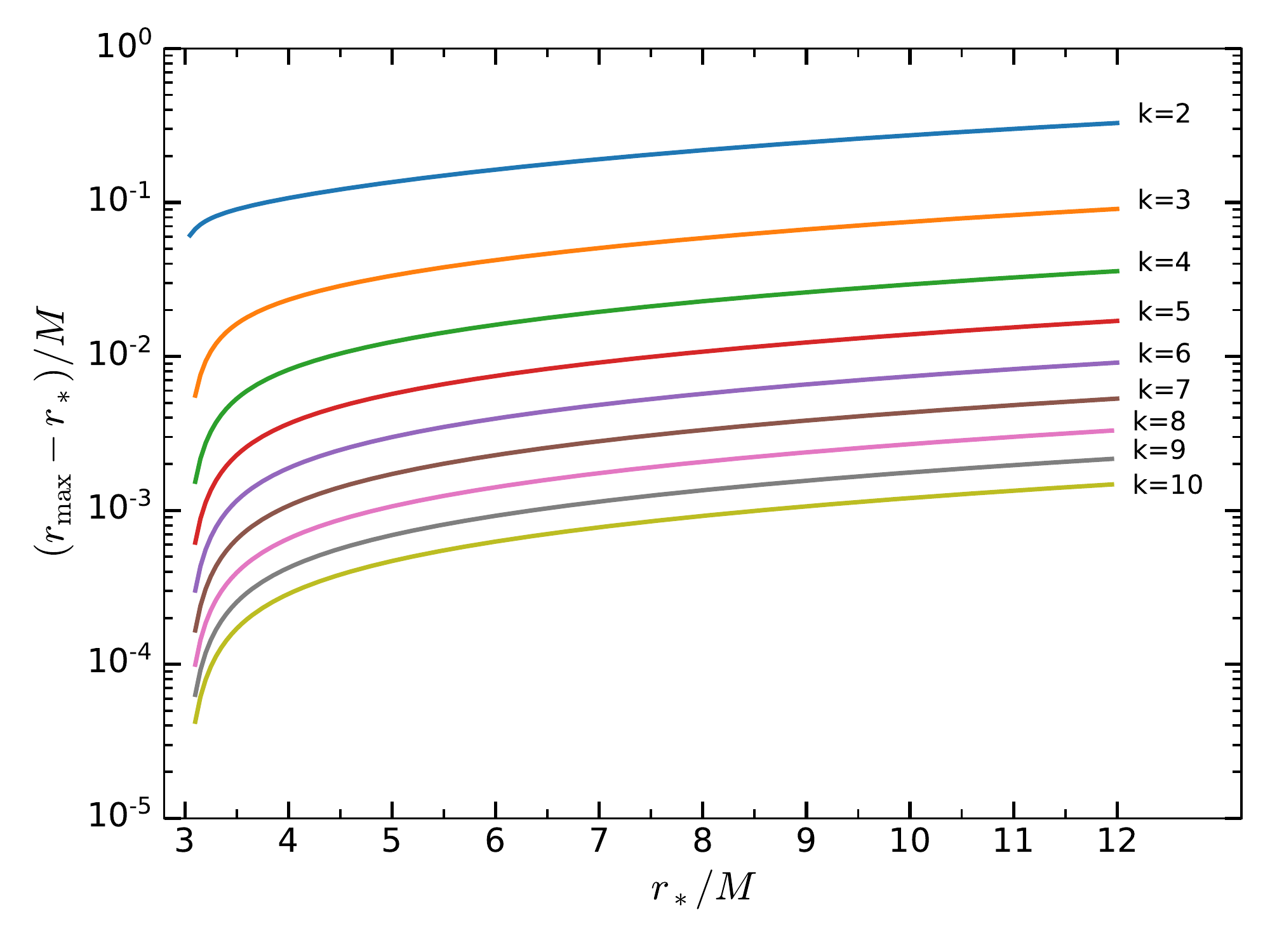}
\caption{Distance between $r_{\rm max}$ location and the stellar surface for first few underdamped modes.}
\label{rstarm_r0m_rel}
\end{figure}

\section{Characteristic frequency maximum: Mass and radius determination}
\label{Section4}
While the frequency of the oscillations in general decreases with luminosity, a distinctive feature to be noted in Figure~\ref{fig:eigefr} is the maximum in the frequencies of underdamped oscillations close to the stellar surface. For given stellar parameters, $M$, $r_*$, we mark the radius at which the atmosphere can oscillate with maximum frequency, $\nu_{\rm max}$, as  $r_{\rm max}$. The maximum for the second mode is clearly pronounced. For higher modes, since $r_{\rm max}$ is quite close to the stellar radius, it would be difficult to observe the decrease of frequency to the left of the maximum as the luminosity decreases. The presence of the frequency maximum can be identified as due to the steep radial decrease of damping close to the star which is mostly attributed to a rapid decrease in viewing angle, as remarked in Section~\ref{damped_oscill_sec},
  in the discussion of eq.~(\ref{blowsup}).

\begin{figure}
\includegraphics[scale=0.43]{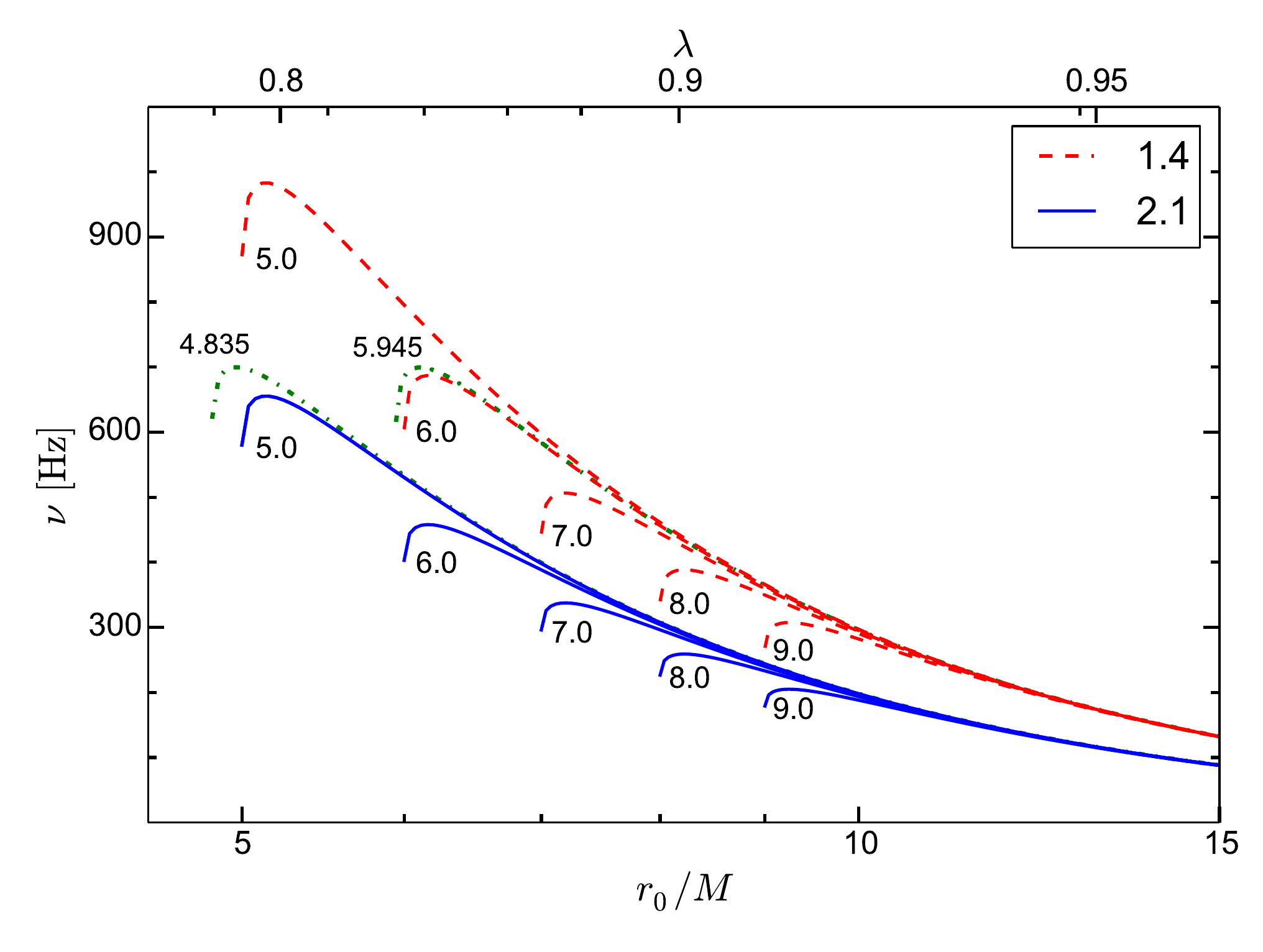}
\caption{Behaviour of the second mode frequency maximum as function of $r_0$ for different stellar masses and radii. Dashed (red) curves are for $1.4\,{\rm M}_{\odot}$ and solid (blue) curves for $2.1\,{\rm M}_{\odot}$. Dash-dotted (green) lines represent the curves for both these masses with frequency maximum value of $700$~Hz. For given mass, various stellar radii ($r_*/M$) chosen are marked for the curves in the figure. }
\label{freq_rstarm}
\end{figure}

Figure~\ref{rstarm_r0m_rel} shows for the first few underdamped modes the coordinate distance between the location of the atmosphere that corresponds to the maximum frequency and the stellar surface. Lower modes are comparatively farther away from the stellar surface, while the distance also increases with the stellar radius. 
\begin{figure*}
    \centering
    \begin{subfigure}[t]{0.5\textwidth}
        \centering
        \includegraphics[scale=0.43]{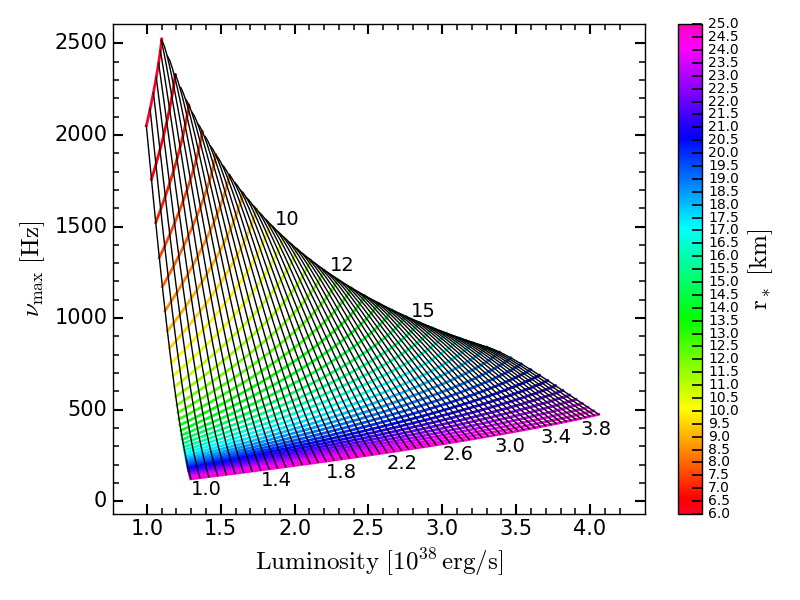}
       \caption{$k=2$}
       \label{obs_freq_lum_2}
    \end{subfigure}%
    \begin{subfigure}[t]{0.5\textwidth}
        \centering
        \includegraphics[scale=0.43]{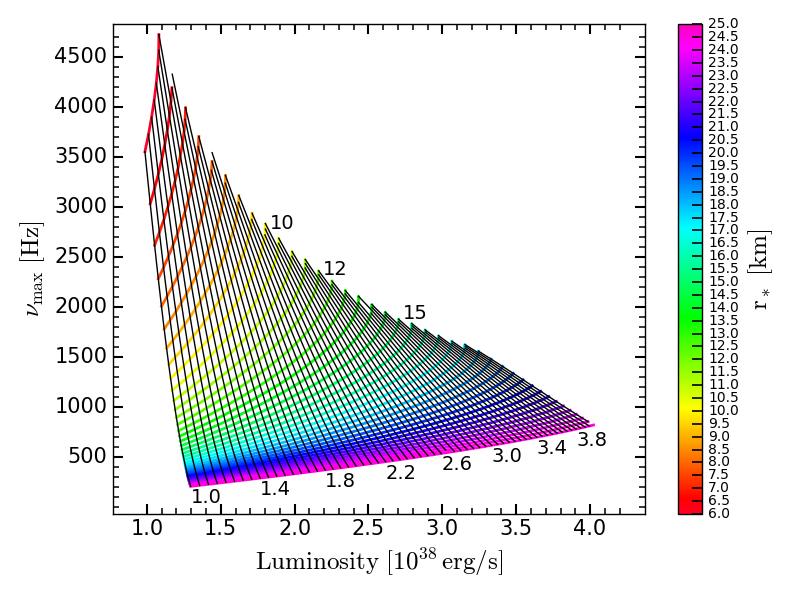}
      \caption{$k=3$}
      \label{obs_freq_lum_3}
    \end{subfigure}%
\caption{Mass and radius of neutron star for given  luminosity and frequency of oscillations for the second and third modes. The lines sloping upwards are the contours of stellar radius in km, plotted in increasing steps of 0.5 km from top to bottom. The nearly vertical black lines are the contours of stellar mass, plotted in increasing steps of $0.05~M_{\odot}$ from left to right.}
\end{figure*}

For a particular mode, this maximum of the frequency, $\nu_{\rm max}$, depends on two parameters ($r_*,~M$). In Figure~\ref{freq_rstarm}, we plot the frequencies of the second mode for $1.4\,{\rm M}_{\odot}$ and $2.1\,{\rm M}_{\odot}$ stellar mass and various stellar radii. The following can be noted from the figure:\\ 
a) For a given stellar mass, $\nu_{\rm max}$ decreases with increasing $r_*/M$. \\
b) For a given $r_*/M$, $\nu_{\rm max}$ is inversely proportional to the stellar mass, as expected for GR scaling. \\
c) Irrespective of the values of $r_*/M$ and $M$, $r_{\rm max}$ is always located close to the stellar surface.\\ 
d) There is a degeneracy in the frequency maximum value that occurs for different combinations of $r_*/M$, $r_{\rm max}/M$ and $M$. For example, 
two different stars with masses $1.4\,{\rm M}_{\odot}$ and $2.1\,{\rm M}_{\odot}$ and radii $5.945 \,M$ and $4.835 \,M$ exhibit the same frequency maximum value of $700$ Hz when their respective locations of the atmosphere are at $6.1\,M$ and $4.95\,M$. These frequencies are plotted as green dash-dotted lines in Figure~\ref{freq_rstarm}. As we shall discuss below, the stellar luminosity breaks this degeneracy and allows us to obtain the stellar parameters unique to the corresponding frequency maximum.\\
e) However, for a given $r_*/M$, the frequency maximum occurs at the same radius (in units of $M$), independent of the stellar mass. This follows immediately from equation~(\ref{freq_damp}),  which is a function of $r_*/M$ and $r_0/M$ alone, once the $1/M$ dependence of $\omega_r$ is taken out.
So, the condition for the maximum, i.e., the first derivative of the frequency with respect to $r_0/M$ being set to zero, gives a relation just between the two radii $r_*/M$ and $r_{\rm max}/M$. 


Given the frequency maximum value and the corresponding luminosity, we can determine the mass and radius of the neutron star. Indeed, we already know that for any single mode exhibiting a frequency maximum, we can now find the value of the dimensionless stellar radius, $r_*/ M$, if we know $r_{\rm max}/ M$. We require one last relation to close the system of equations that enable us to solve for the two parameters ($ r_*, \, M$). Equation~(\ref{ECS}) fulfils this requirement by providing a relation between $M$ and $r_{\max}$ through $\lambda$ or stellar luminosity.

  To derive the stellar parameters, the set of nonlinear relations are solved semi-analytically using the following protocol:\\
 i)~We derive $\lambda$ as a function of $M$ and $L_\infty$ from equation~(\ref{lambda}), assuming the Eddington limit for ionized hydrogen. An atmosphere with a different composition would have a different Eddington limit and the $x$-axis of the plots in Figure~\ref{obs_freq_lum_2} and \ref{obs_freq_lum_3} would then vary accordingly.\\
 ii)~With the direct dependence on $\lambda$, $r_{\rm max}/M$ is then   determined as a function of $M$ and $L_\infty$, using equation~(\ref{r_ecs}) with $r_0=r_{\rm max}$.\\
 iii)~For the assumed mode number (say~$k=2$ or $k=3$), we set the first derivative of frequency with respect to $r_0$ to zero to derive $r_*/M$ as a function of $r_{\rm max}/M$, and therefore as a function of $M$ and $L_\infty$.\\
 iv)~The final equation for the frequency maximum is thus solely a function of mass and luminosity, so that given any two quantities, the other can be determined. Thus, given that we know the observed frequency maximum and luminosity, we can determine the mass of the star.\\
 v)~Knowing $M$ and $r_*/M$ we determine the stellar radius.
 
We admit a source of ambiguity in the measurements of radius and mass due to the mode number. 
The mode number is a discrete parameter, and so assuming different mode numbers introduces large
discrete changes in the measured mass and radius.
We expect the higher modes to be less likely to be excited than the breathing mode ($k=2$), given their more complicated eigenfunctions. However, in any case, we believe it would in principle be possible to determine the mode number by observing the frequency variation with varying flux (luminosity)~--~the higher the mode number, the more rapid the variation, as is clear from Figure~\ref{fig:eigefr}.

In Figures~\ref{obs_freq_lum_2} and \ref{obs_freq_lum_3}, we show a grid of stellar mass and radius contours for a given range of luminosity and frequency maximum computed as described above for the second and third modes, respectively. The coloured contours (sloping upwards with luminosity) are for stellar radius in increasing steps of 0.5~km from top to bottom. The nearly vertical black contours represent mass in range of $1\,{\rm M}_{\odot}$ to $4\,{\rm M}_{\odot}$ in increasing steps of $0.05\,{\rm M}_{\odot}$ from left to right. At a given luminosity, $\nu_{\rm max}$ of a star increases with its compactness. Since we assumed that the stellar surface lies above the photon orbit, we omit solutions with $r_*<3M$. Hence, for combinations of higher luminosity and higher frequency maximum, we see no contours in the figures.

\begin{figure*}
    \centering
    \begin{subfigure}[t]{0.5\textwidth}
        \centering
        \includegraphics[scale=0.43]{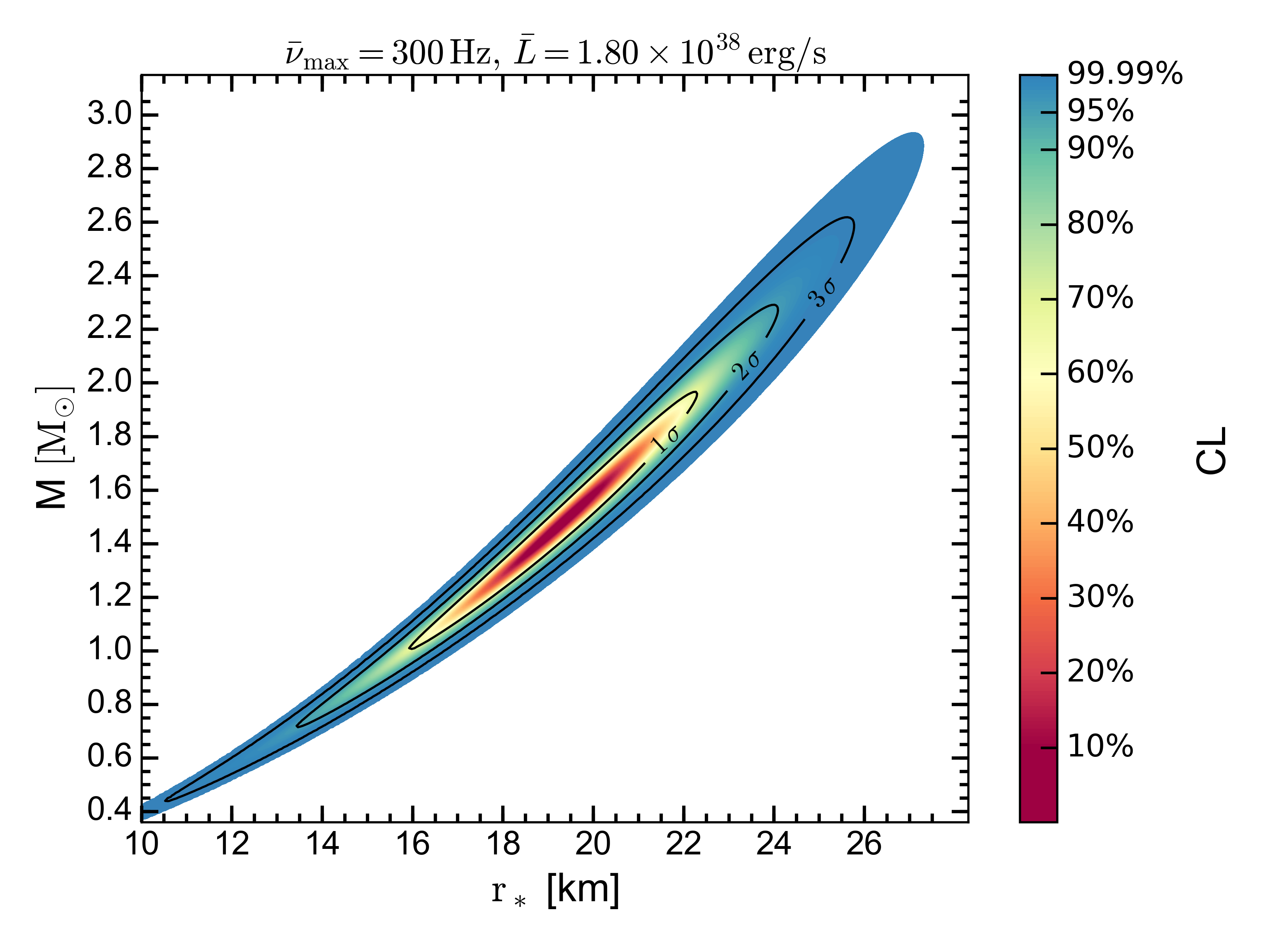}
       \label{chi3001p8}
    \end{subfigure}%
         \begin{subfigure}[t]{0.5\textwidth}
        \centering    
              \includegraphics[scale=0.43]{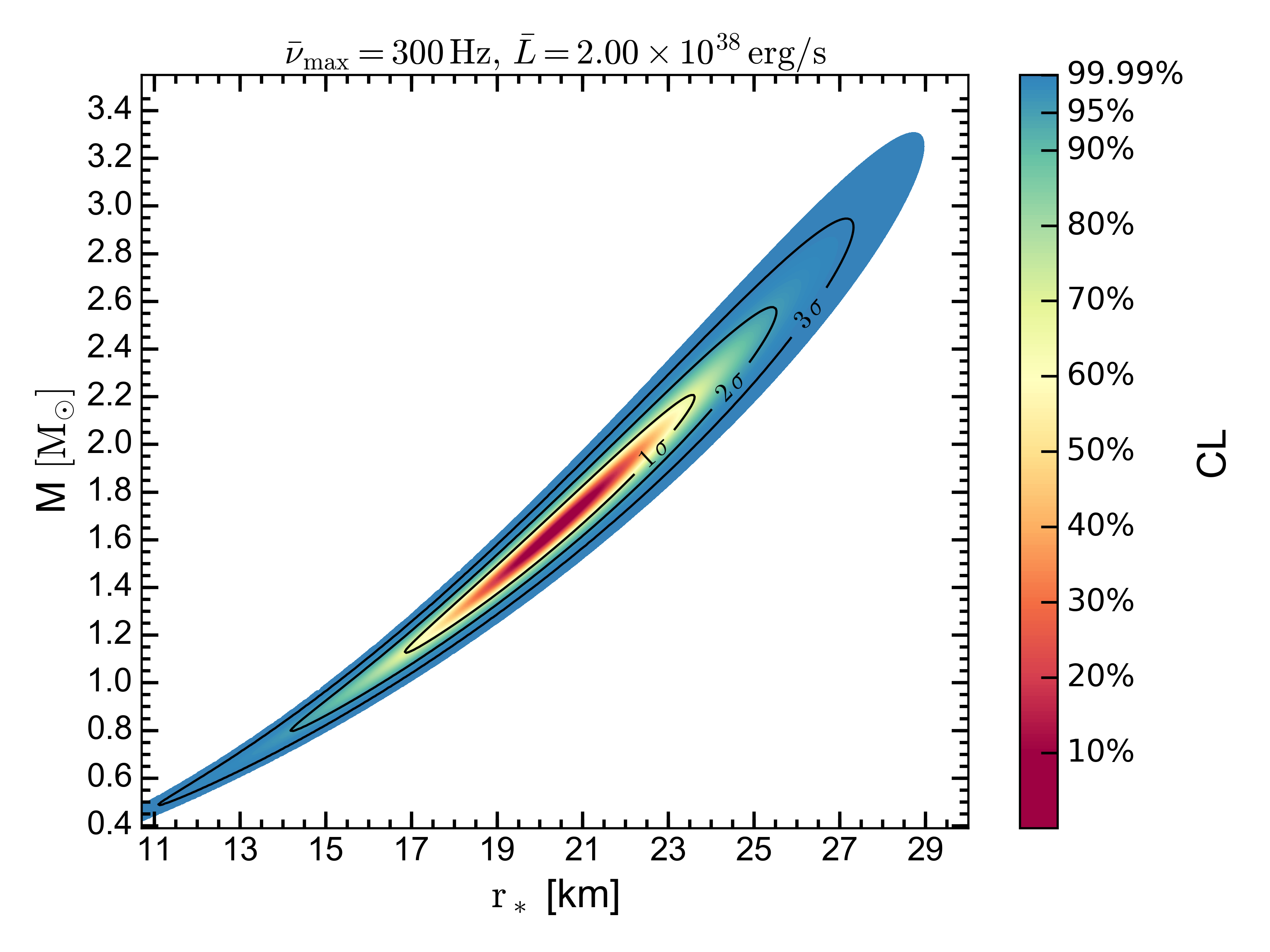}    
       \label{chi3002}
    \end{subfigure}%
    \newline
    \begin{subfigure}[t]{0.5\textwidth}
        \centering
        \includegraphics[scale=0.43]{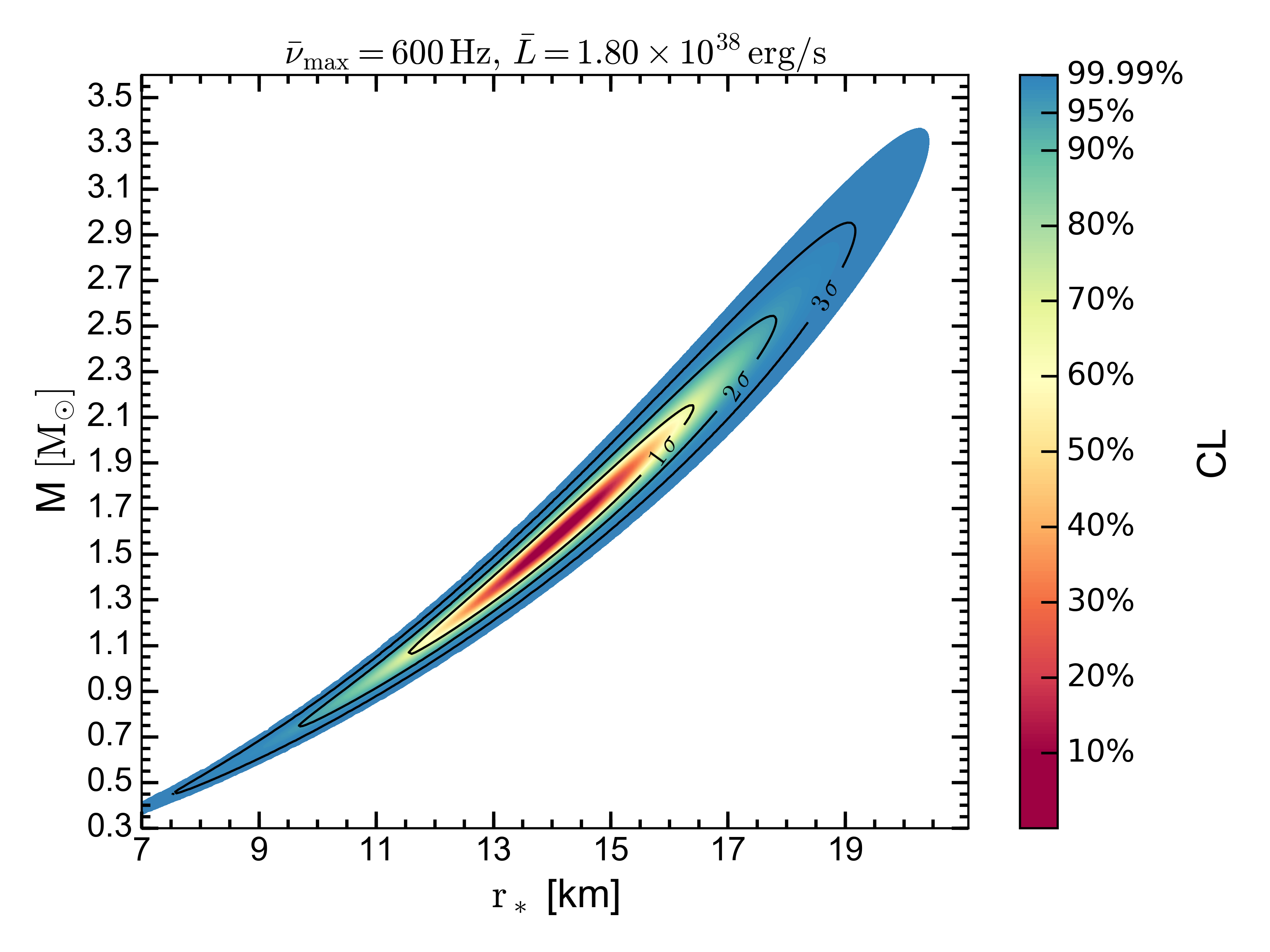}
      \label{chi6001p8}
    \end{subfigure}%
        \begin{subfigure}[t]{0.5\textwidth}
        \centering
        \includegraphics[scale=0.43]{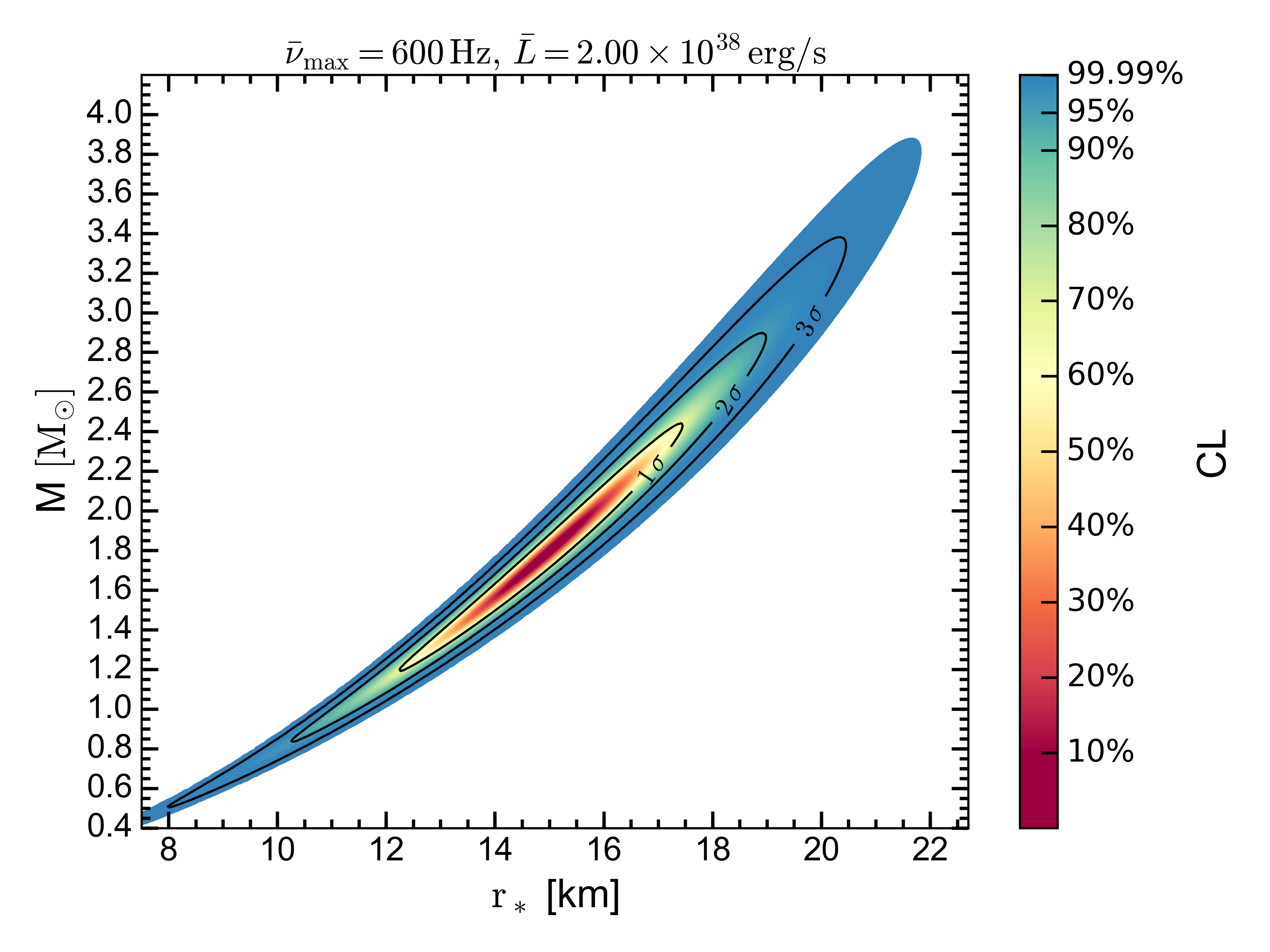}
      \label{chi6002}
    \end{subfigure}%
    \newline
           \begin{subfigure}[t]{0.5\textwidth}
        \centering
        \includegraphics[scale=0.43]{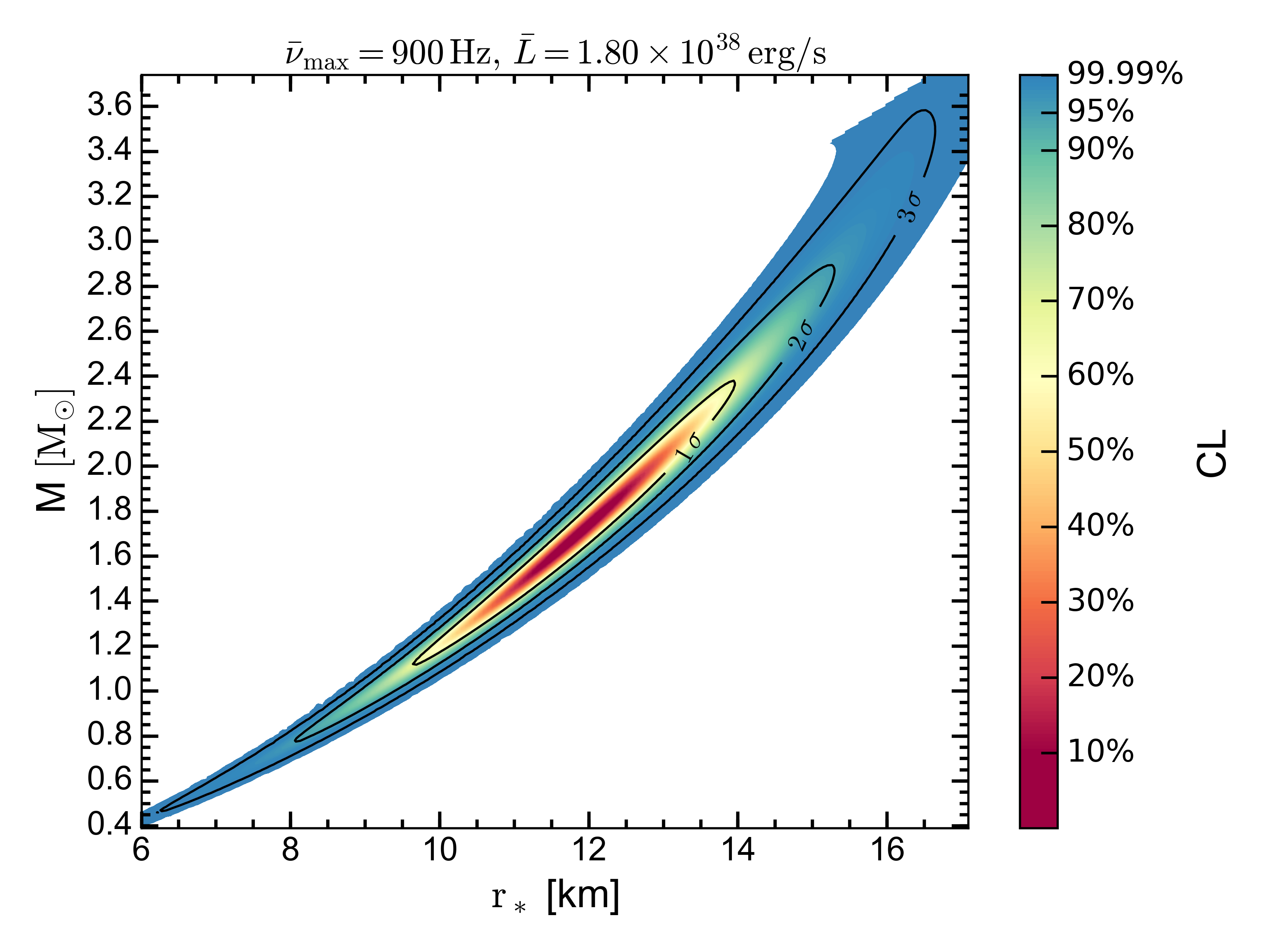}
      \label{chi9001p8}
    \end{subfigure}%
        \begin{subfigure}[t]{0.5\textwidth}
        \centering
        \includegraphics[scale=0.43]{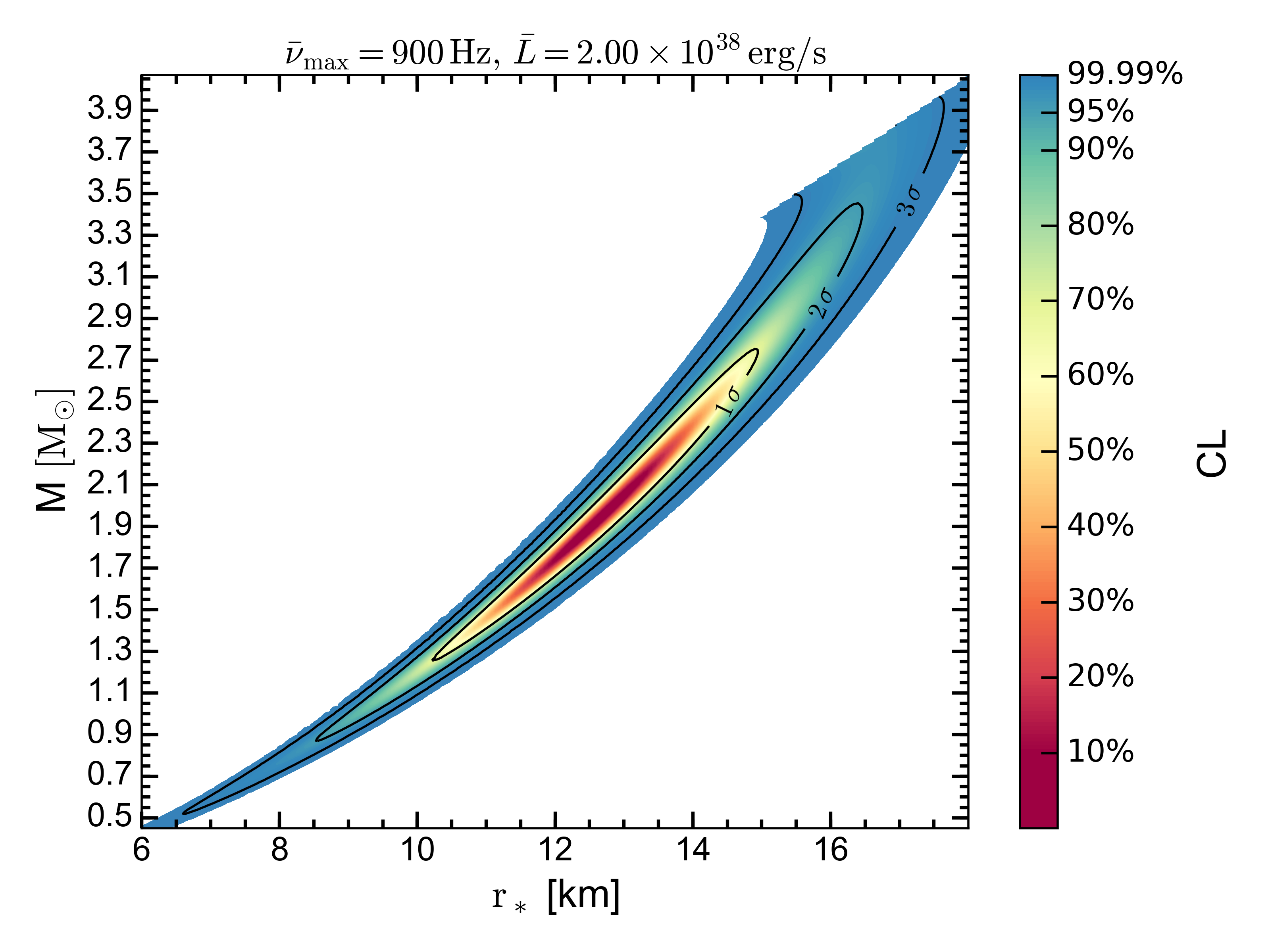}
      \label{chi9002}
    \end{subfigure}%
\caption{Contours of confidence levels (CL) for the mass and radius of the neutron star are shown corresponding to different mean values of frequency maximum $\bar{\nu}_{\rm max}$ and luminosity ($\bar{L}$) as labelled in the titles of the plots. In all the plots, the errors on frequency maximum and luminosity are taken to be $3~\%$ and $20~\%$ respectively. The $1\sigma$, $2\sigma$ and $3\sigma$ error contours are overplotted in black.}
\label{banana}
\end{figure*}
As we can note from the grid in the plots, each pair of stellar luminosity and frequency maximum thus gives a unique measure of the radius and mass of the neutron star. For example, if the observed frequency maximum is around 600~Hz and the corresponding luminosity of the star observed at this frequency is around $1.9\times10^{38}~\mathrm{erg~s}^{-1}$, then assuming that the frequency corresponds to the second mode ($k=2$), the determined stellar radius is $14.5$~km and mass is $1.65~{\rm M}_{\odot}$. If we assume the third mode ($k=3$), then the determined stellar radius increases to $18.5$~km and the mass estimate decreases to $1.6~{\rm M}_{\odot}$. 

Inferring the mass and radius of the neutron star through this method thus requires only a single observation of two quantities, luminosity and frequency. However, this method is subject to certain systematic uncertainties, of which the most important is the uncertainty in the distance to the source, which translates directly into uncertainty in the luminosity, thereby leading to errors in the estimated mass and radius. We discuss these in the next subsection.

\subsection{Error budget}
The determination of the neutron star mass and radius, described in this section, would be quite accurate if the distance to the source were known quite precisely, as in the 47 Tuc globular cluster, with quoted distance $d=4.53\pm 0.06\,$kpc \citep{woodley+12}. Unfortunately, the distance  errors to other globular clusters are less constrained, with uncertainties sometimes approaching $10\%$~\citep{guillot+13, steiner+18}. Typical errors on the distances to the neutron star LMXBs in the field are even larger, falling between $10-20\%$~\citep{galloway+08}. Such large uncertainties in the distance imply that the typical uncertainties in luminosity can be around $20\%$ or higher.
  
Frequencies of oscillations, once detected, are typically much more precise. Planned, sensitive X-ray instruments e.g. \textit{STROBE-X} \citep{strobex} and \textit{eXTP} \citep{extp}, should be able to measure the frequencies to exquisite precision, given by the ratio of the oscillation period to the time-scale for frequency change (as the luminosity varies), which should be less than one part in a hundred. To be conservative, in the remainder of this section we assume a frequency error of $3\%$. All the analysis throughout the rest of this paper considers only the second mode, i.e., $k=2$.

To quantify the mass and radius errors, we do a simple $\chi^2$-error analysis with chosen uncertainties in luminosities and frequencies. We assume that the luminosity and frequency maximum measurements are normally distributed, with different mean values for different cases. We consider the systematic uncertainties on the luminosity and frequency to be $20$ per cent and $3$ per cent respectively, which are taken to be the standard deviations of the corresponding distributions. We consider six cases in total with the mean luminosities ($\bar{L}$) to be $1.8\times10^{38}$ and $2.0\times10^{38}~{\rm erg~s}^{-1}$, and the mean frequency maxima ($\bar{\nu}_{\rm max}$) to be 300, 600, and 900 Hz. Figure~\ref{banana} shows the joint confidence regions for mass and radius for the six different cases with $\bar{L}$ and $\bar{\nu}_{\rm max}$ given in the title of the each plot. The confidence regions are determined based on the $\Delta\chi^2$ values calculated with the chosen uncertainties. Overplotted black contours denote the $1\sigma$, $2\sigma$, and $3\sigma$ errors that correspond to the $68.27\%$, $95.45\%$, and $99.73\%$ confidence levels, respectively.

A projection of the $1\sigma$ contour ($\Delta \chi^2 =1$) on to the corresponding axes yields upper limits to the $1\sigma$ errors on mass and radius, but the actual $1\sigma$ errors on the parameters are always smaller.  As we cannot assert that these parameters have a Gaussian distribution, we turn to a Monte Carlo analysis to generate the probability distributions for mass and radius and to accurately determine the errors on their values.

\begin{figure}
\includegraphics[scale=0.35]{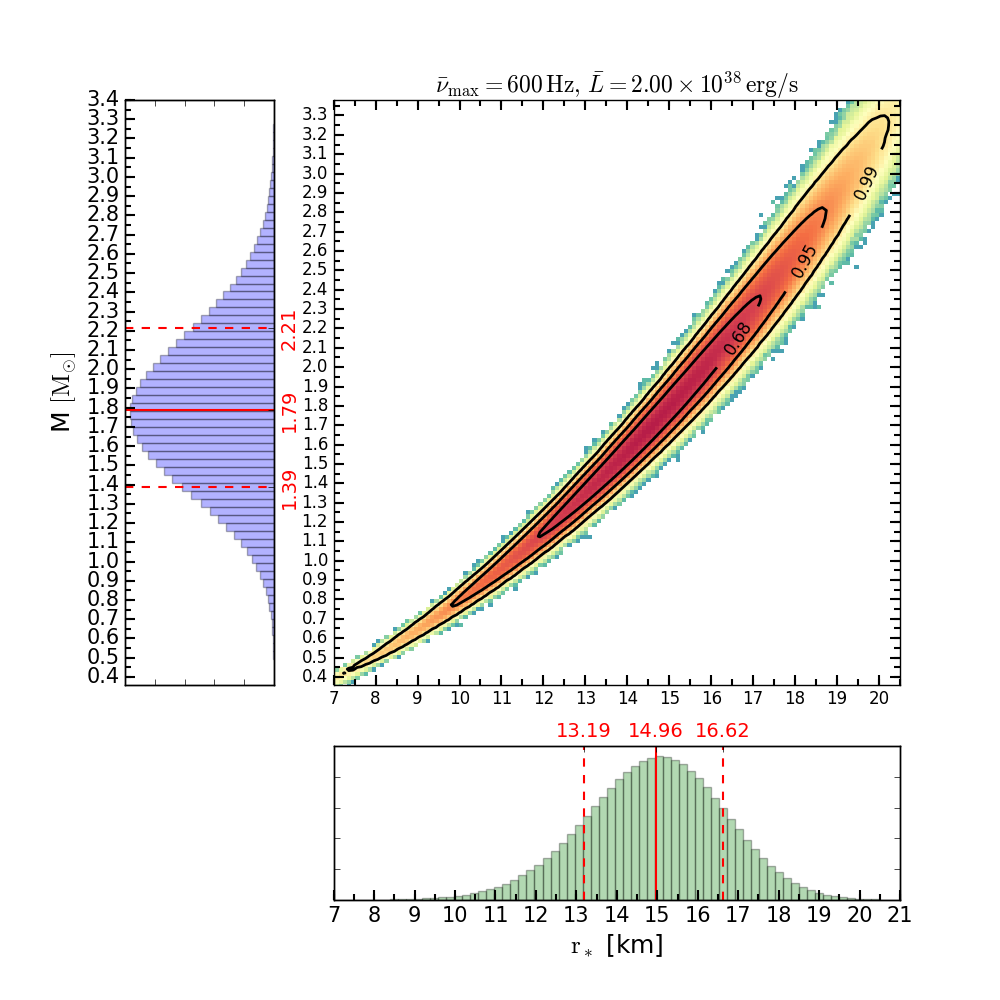}
\caption{Probability distributions and mass and radius computed though the Monte Carlo simulations. The median and the $1\sigma$ range for each quantity are marked with solid-red and dashed-red lines. The central plot shows the joint confidence region for the two quantities, with the $68\%$, $95\%$, and $99\%$ confidence levels represented by the black contours.}
\label{hist_600_2}
\end{figure}

We once again consider the six cases with the assumption that the frequency and luminosity are normally distributed with the mean values as given in the first two columns of Table~\ref{table1} and with the same $3\%$ and $20\%$ uncertainties as before. For each such case, we perform a Monte Carlo simulation to generate a large ensemble of $10^8$ pairs of luminosity and frequency maximum values that are normally distributed with the chosen $\bar{\nu}_{\rm max}$ and $\bar{L}$. For each such pair of luminosity and frequency maximum, we compute the corresponding pair of mass and radius to generate the posterior distributions for the mass and radius. In Figure~\ref{hist_600_2} we show the distributions for mass and radius for one such case ($\bar{\nu}_{\max} = 600$ Hz and $\bar{L}=2.0\times10^{38}~{\rm erg~s}^{-1}$). The red solid line represents the median (50th percentile), while the red dashed lines denote $1\sigma$ error bounds (16th and 84th percentile). The central plot shows the contours of joint confidence regions at the levels 0.68 and 0.95. The $1\sigma$ errors on mass and radius determined through this method are given in the third and fourth columns of Table~\ref{table1}. The typical errors on mass and radius from this table are slightly larger than $20$ per cent and $10$ per cent, respectively. 

\begin{table}
\begin{center}
\begin{tabular}{ c c c c }
 \hline
$\bar{\nu}_{\rm max}$ (Hz) & $\bar{L}$ ($\times 10^{38}$ ${\rm erg~s}^{-1}$) &$M$ (${\rm M}_{\odot}$) & $r_*$ (km)\\
 \hline
 \vspace{2mm}
300& 1.8 & $1.48^{+0.32}_{-0.32}$& $19.3 ^{+2.0}_{-2.2}$ \\
 \vspace{2mm}
300& 2.0 & $1.65^{+0.37}_{-0.35}$&  $20.4^{+2.1}_{-2.3}$ \\
 \vspace{2mm}
600& 1.8 & $1.58^{+0.35}_{-0.32}$&  $14.1^{+1.5}_{-1.6}$ \\
 \vspace{2mm}
600& 2.0 & $1.79^{+0.42}_{-0.40}$&  $15.0^{+1.7}_{-1.8}$ \\
 \vspace{2mm}
900& 1.8 & $1.70^{+0.41}_{-0.37}$&  $11.9^{+1.3}_{-1.4}$ \\
 \vspace{2mm}
900& 2.0 & $1.94^{+0.48}_{-0.44}$&  $12.6^{+1.4}_{-1.5}$ \\
\hline
\end{tabular}
\end{center}
\caption{Frequency maximum and luminosity are assumed to be normally distributed with the mean values as given in the first and second columns, respectively. Corresponding $1\sigma$ errors, which are given in the third and fourth columns, are determined from the probability distributions of mass and radius.}
\label{table1}
\end{table}

\begin{figure*}
\centering
\begin{subfigure}[t]{0.5\textwidth}
\centering
\includegraphics[scale=0.43]{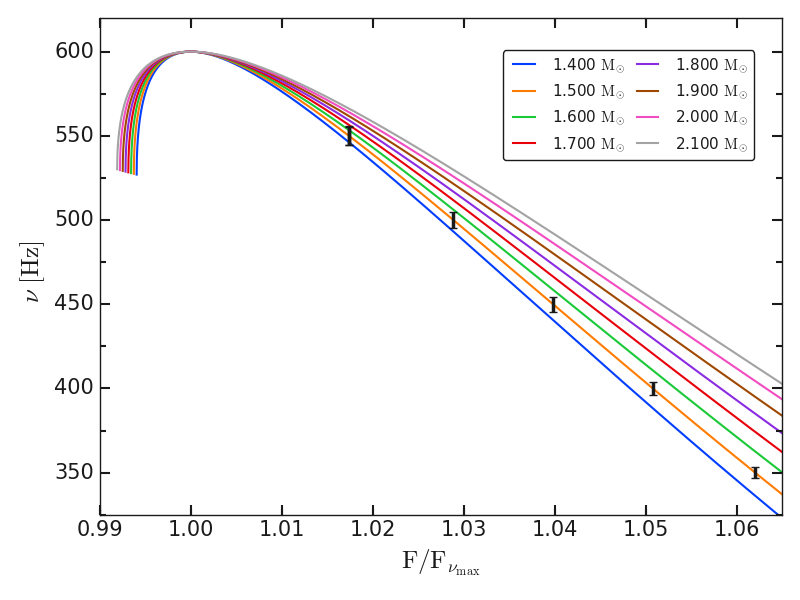}
\caption{ }
\label{ffr600}
\end{subfigure}%
\begin{subfigure}[t]{0.5\textwidth}
\centering
\includegraphics[scale=0.43]{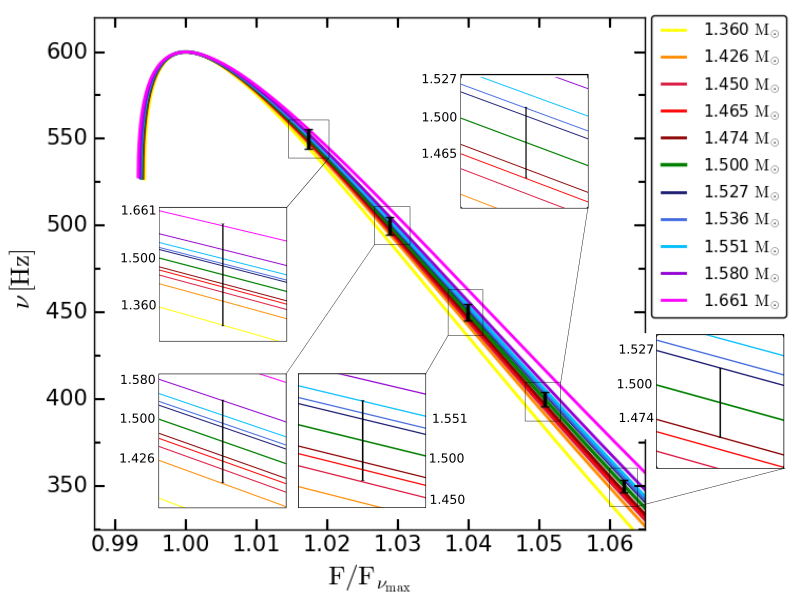}
\caption{}
\label{error_ffr600}
\end{subfigure}%
\caption{Frequency variation with flux for different masses labelled in the plot that have the same frequency maximum value at $600$~Hz. In the left-hand panel (a), the increasing order of the curves (from the lowest curve in blue to the highest curve in grey) have their the corresponding masses increasing in steps of $0.1~{\rm M}_{\odot}$ and the respective radii are $r_*=~13.26,~13.72,~14.164,~14.594,~15.011,~15.416,~15.809,~16.192$ km. In the right-hand panel (b), the masses labelled in increasing order correspond to the increasing order of the curves with the respective radii given by $13.073,\,13.38,\,13.493,\,13.559,\,13.601,\,13.72,\,13.841,\,13.881,\,13.948,\,14.076,\,14.427$~km.}
\label{fits}
\end{figure*}

Therefore, we can say that with respective errors of $20\%$ and $3\%$ in luminosity and frequency, we can estimate the parameters of the neutron star with an accuracy of $20\%$ for mass and $10\%$ for radius. It is easy to see that the errors on mass and radius will be significantly lowered if the errors on luminosity can be decreased. Indeed, the uncertainty in the determined values of mass and radius are of the order of the uncertainty in the luminosity.

Interestingly, if the oscillation frequency is observed over a wide range of fluxes, the neutron star mass and radius could be determined without prior knowledge of the distance to the source. In the following section, we discuss how to simultaneously determine the mass, radius, and distance to the source, and show that the errors on $M$, and $r_*$ would be decreased by an order of magnitude relative to the errors discussed above.

\section{Mass, radius, and distance determination}
\label{Section5}
One way to circumvent the problem with distance uncertainties is to consider the ratio of luminosities that essentially translates into flux ratio, which is a directly measurable quantity like frequency. The two green dashed-dotted lines in Figure~\ref{freq_rstarm} that share the same frequency maximum for different masses and radii show visibly different frequency variation with $\lambda$ ($\propto~L/M$), but the mass, and therefore $\lambda$ as well, is not yet known, so in practice we cannot use this plot.
In fact, let us assume that we do not know the distance to the source, so that we do not know the luminosity.
 None the less, if we consider plotting the frequency variation with the flux ratio, which does not depend on the distance, we can distinguish the curves for different masses and radii. Suppose that we have a set of observed data points for frequency variation with flux. In principle, as per our model, there exists a unique pair of mass and radius, within error, that can fit all the observed data points. Although, in reality, if we do not have a sufficient number of data points, or data spanning a sufficiently wide range over frequencies, then a wide range of masses and radii can potentially fit the observed data points, thus making it difficult to stringently constrain the mass and radius (unless we already know the luminosity, in which case we can proceed as in Section~\ref{Section4}). 

  To elaborate on this, we consider an example of frequency variation with luminosity for the parameters $M = 1.5~M_{\odot}$ and $r_* = 13.72$ km. The curve has $\nu_{\rm max}~=$~600~Hz and we chose five data points on this frequency curve corresponding to 550, 500, 450, 400 and 350 Hz with $1\%$ uncertainty. We can use these five data points as a proxy for the real observation data set and check to within what accuracy we can recalculate the values of mass and radius by finding the best fits to these points. Since we do not know the distance, these points can only be represented on the frequency-flux plane if we scale the corresponding luminosities of each point with the luminosity of any one particular point. This is the same as scaling the fluxes of the observed data points to the flux of one data point. In this example, we chose this reference point to be the one with frequency 600 Hz.

  First, let us suppose that we have a precise measurement for the frequency maximum. In such a case, we can try to fit through all the points with the frequency curves for various mass and radius pairs that have the same $\nu_{\rm max}~=~600$~Hz, as shown in Figure~\ref{ffr600}. The $x$-axis for each curve is scaled with its corresponding flux (${\rm F}_{\nu_{\rm max}}$) at $\nu_{\rm max}$. The lower we go in the frequency  data compared to $\nu_{\rm max}$, the better we can constrain the mass and radius measurements. An inspection of the plot close to the error bars located at $400$~Hz and $350$~Hz shows that we can already limit the errors on mass to a much better accuracy than $\pm~0.1~{\rm M}_{\odot}$. This can be better visualized in Figure~\ref{error_ffr600}, which shows fits to the error bounds for each data point. This way the errors on mass and radius determined from the fits to the bounds of the 350 Hz error bar are a factor of $6$ less compared to the errors from the fits to the 550 Hz error bar. So given a precise measurement for $\nu_{\rm max}$, and a wide range of frequency measurements, the mass and radius can be estimated with relatively minor errors of $< 2\%$ ($\pm 0.027\,{\rm M}_{\odot}$) and $< 1\%$ ($\pm 0.121$ km) respectively.
  
\begin{figure}
\includegraphics[scale=0.43]{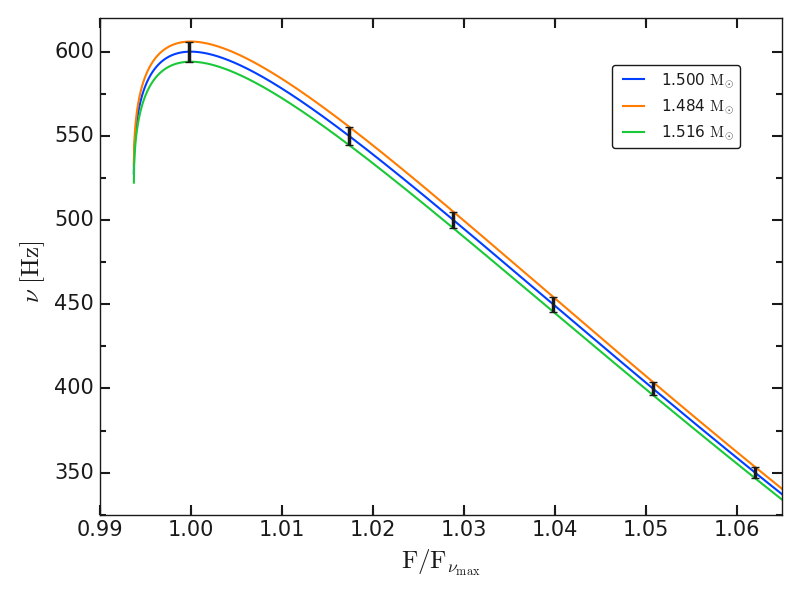}
\caption{Frequency curve fits when the frequency maximum is also measured with an uncertainty. The respective radii for the curves with masses $1.5,~1.484,~1.516~{\rm M}_{\odot}$ are $r_*~=~13.72,~13.579,~13.862$~km. }
\label{error_600}
\end{figure}

  Another potential source of error is from a measurement of the frequency maximum with given uncertainty just like the remaining data points. We fit three curves to the centre, lower end, and upper end of the error bars as shown in Figure~\ref{error_600}. The respective masses are labelled in the plot, and we see all the curves passing through the ``data'' points in Figs.~\ref{fits} and \ref{error_600} yield values of $M$ and $r_*$ within an error of less than 2\%.

   Thus, we conclude that the overall error in the neutron star mass and radius determination by this method is of the same order as the assumed error in the measured frequencies ($~1$\%), which would be an excellent result.

\subsection{A note on distance estimates}

\begin{figure}
\includegraphics[scale=0.43]{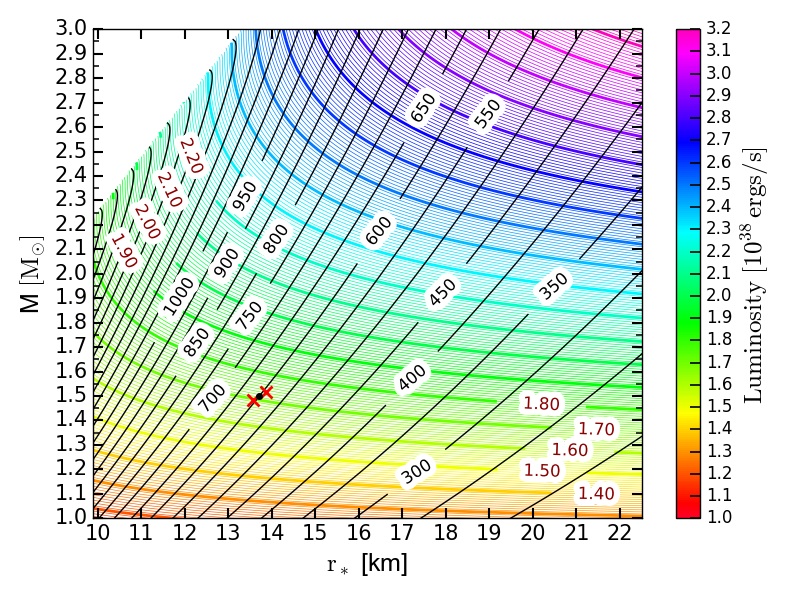}
\caption{A grid of luminosity and frequency maximum contours computed from a wide range of masses and radii. The coloured downward sloping contours represent luminosity (in $10^{38}~{\rm erg~s}^{-1}$ units) plotted in steps of $0.1$, and the black quasi-diagonal  lines represent the frequency maximum contours plotted for every 50 Hz. The errors on mass and radius determined in Section 5 correspond to the interval (close to the 600 Hz line) marked by the red crosses.}
\label{rstar_mgrid}
\end{figure}

With the above derived constraints on mass and radius for a known $\nu_{\rm max}$, we can determine the luminosity and thereby the distance to the source with a high accuracy. In Figure~\ref{rstar_mgrid}, we show a grid of $\nu_{\rm max}$ (nearly diagonal lines in black) and luminosity contours (coloured downward sloping lines) plotted for a wide range of masses and radii. The red crosses lying close to the intersection of $\nu_{\rm max}=600$~Hz and $L=1.70\times 10^{38}\,{\rm erg~s}^{-1}$ contours mark the mass and radius errors estimated with $1\%$ uncertainty in $\nu_{\rm max}$, as just described, and with these, the luminosity can be determined with an accuracy of $1\%$. This implies that the distance to the source can also be constrained to within $1\%$ errors.

The technique may have astrophysical applications. Type-I X-ray bursts with photospheric radius expansion have been treated as standard candles to evaluate distances to globular clusters containing these X-ray bursters with uncertainties on the level of $30\%$ \citep{kuulkers+03}. Measurements of LMXBs in quiescence seem to be more precise, with errors less than $10\%$~\citep{guillot+13, steiner+18}. With our model, if it is successfully applied to X-ray bursters, we could place much more stringent constraints on the distances to these globular clusters. For clusters such as 47 Tuc, where the distance has been estimated accurately through other methods \citep{woodley+12}, one would get a valuable cross-check on the validity of our approach. 

%

\section{Further discussion}
\label{SectDisc}
We have derived the modes of oscillation of an atmosphere suspended above the neutron star surface by the force of radiation. Although the fundamental mode is critically damped due to radiation drag, the higher oscillatory modes of optically thin atmospheres may be detected by future X-ray missions. We have shown that with the knowledge of the frequency and the corresponding near-Eddington luminosity from the X-ray observations, the mass and radius of the neutron star can be derived. It would be interesting to check the applicability of this method with the currently available X-ray data, to compare with the already available estimates of the neutron star mass and radius.

Recent works on constraining the mass and radius of neutron stars rely on X-ray spectroscopy of the thermal emission from the hot neutron stars residing in quiescent LMXBs. An analysis of LXMB sources in the Galactic plane gave mass and radius constraints on two sources with respective errors at $90\%$ confidence level between $~30\%$ and $~50\%$~\citep{marino+18}, and comparable, or even higher, errors for several globular cluster sources~\citep{steiner+18}.  These error estimates assume a precisely known distance, the errors further increase when accounting for distance uncertainties. For quiescent LMXBs X7 in the globular cluster 47 Tuc, the distance to which is known accurately, \citet{bogdanov+16} obtains a $7\%$ error in the radius, assuming a particular value for the neutron star mass. Our method, which is described in Section~\ref{Section4}, is potentially much more accurate (and yields both, the mass and radius), but for uncertain distances it yields errors as large as those of the distance determination. However, we note that the slope (in the mass--radius diagram) of the confidence level contours differs between the methods. The spectral technique employed to various LMXB sources in globular clusters yields mass and radius constraints with probabilities shifting towards larger radii for lower masses in most of the sources~\citep{steiner+18}. This picture when overlapped with Figure~\ref{banana} of this paper in which the probabilities shift towards larger radii for larger masses can further narrow down the limits on mass and radius. Thus in an overall comparison, burst oscillation study can be a very useful technique to constrain the mass and radius with a high accuracy.

As we showed in Section~\ref{Section5}, if the oscillation is detected over a sufficiently wide range of X-ray fluxes, the distance to the source may be directly determined by comparing observations with the theoretical frequency--flux curves. In such a case, the mass and radius, as well as the distance, could be determined to high accuracy with no reference to the spectral method.
 
Apart from the oscillations and their relevance to the observed frequencies, the levitating atmosphere may have its own observational signatures. A study by \citet{Rogers2017} suggests that such levitating atmospheres can deflect the light rays coming from the central compact source and this can significantly affect the appearance of the central object in the observations.

Similar atmospheres are found in the corona of accretion discs and described as photon floaters, formed at the critical height where the radiation force from the disc is balanced by the gravitational force of the central object \citep{fukue96}. It would be interesting to see if similar oscillations persist in the disc corona. Under the plane-parallel approximation, the eigenfunctions of these atmospheric oscillations may remain the same as for the spherically symmetric geometry presented here, while the frequencies of such oscillations would be different. 

So-called burst oscillation frequencies are detected during the Type I X-ray bursts. If these are identified with the actual oscillations of an optically thin atmosphere, the methods described in this paper can be directly applied to the X-ray burst frequencies in order to determine the neutron star mass and radius.  However, the radiating surface in X-ray bursters is optically thick and throughout this work we have assumed an optically thin fluid.  The construction of optically thick atmospheres was given by \citet{Wielgus2016} and the procedure involved numerical methods. So an analytical analysis of the oscillation modes in such a case is non-viable  and their analysis is beyond the scope of this paper. Clearly, they are of great astrophysical interest and could possibly be subject to less radiation drag since radiation transport inside the optically thick atmosphere is due to diffusion. It is hoped that future work will treat the oscillations of the optically thick levitating atmospheres, especially in the context of radius expansion X-ray bursts. Regardless of this, future instruments with high time resolution and good sensitivity to higher energy photons, such as \textit{STROBE-X} \citep{strobex} and \textit{eXTP} \citep{extp}, may detect optically thin atmospheres, and their oscillations, outside Eddington luminosity neutron stars.
\section{Conclusion}
\label{SectConc}
Neutron stars with super-Eddington luminosity have a strong radiation field close to the stellar surface that dominates over gravity. The strong radiation force pushes the surrounding matter away from the stellar surface to a certain critical radius $r_0$, beyond which gravity prevails, thereby forming an atmospheric shell at $r_0$ that levitates above the stellar surface. These levitating atmospheres are thus supported by the strong radiation flux from the stellar surface. The same flux is a source of strong radiation drag that eventually damps any oscillations of the atmosphere.

In this paper, we analytically study in general relativity the oscillations of levitating atmospheres that are optically and geometrically thin, including the radiative terms in the perturbation equation. Radiation drag induces damping of oscillations in all regimes and we find that the damping coefficient is independent of the mode number. Radiation drag prevents oscillations only for
the first mode, while higher modes are underdamped. The frequency range observed for the burst oscillations in the decay phase of the X-ray bursts is 300--600 Hz, which lies within the obtained frequencies of the damped oscillations. The frequency of these oscillations increasing with decreasing luminosity is in qualitative agreement with the observations for the oscillations in the decay phase of the X-ray burst where luminosity decreases with time.

The frequency of these oscillations exhibits a characteristic maximum that is more pronounced for the lower order modes. Noting that the frequency is only dependent on the stellar luminosity, mass and radius, we compute the mass and radius as a function of the maximum frequency, the luminosity. However, the accuracy of this method is limited by the uncertainties in distance to the source, typically of $\sim 10\%$, which lead to a comparative relative error on mass and radius estimates.

An alternative method to determine the stellar parameters is also proposed, which does not rely on prior knowledge of the distance. This method focuses on the variation of the two directly observable quantities with respect to each other, i.e., the frequency variation with the flux. We show that with this method, the stellar parameters could be estimated as accurately as the frequencies are measured, i.e., to  $\sim 1\%$ should such oscillations be observed over a sufficiently wide range of frequencies. The new constraints on the mass and radius when combined with the knowledge of the frequency maximum allow us to also directly infer the luminosity of the source with an accuracy of $<1\%$. Therefore, the distance can also be constrained to within a $1\%$ error, by comparing the inferred luminosity with the measured flux. This model thus potentially establishes a new way to determine, with exceptionally high accuracy, the distance to the neutron star, and its mass and radius.

\section*{Acknowledgements}
The authors would like to thank the anonymous referee for the interesting and helpful comments that helped improve the quality of the paper.
This research was partly supported by the Polish NCN grants UMO-2013/08/A/ST9/00795, 2013/10/M/ST9/00729 and in part by the National Science Foundation under Grant No. NSF PHY11-25915. MW acknowledges support by the Black Hole Initiative at Harvard University, through the grant from the John Templeton Foundation.




\bibliographystyle{mnras}
\bibliography{oscil} 








\bsp	
\label{lastpage}
\end{document}